\documentclass[preprints,article,accept,moreauthors,pdftex]{Definitions/mdpi}
\firstpage{1} 
\pubvolume{xx}
\issuenum{1}
\articlenumber{5}
\pubyear{2022}
\copyrightyear{2022}
\history{Received: 08/07/2022, Update: 31/08/2022}

\pdfoutput=1

\usepackage{subcaption}

\Title{Safe reinforcement learning for multi-energy management systems with known constraint functions}

\Author{Glenn Ceusters $^{1,2,3,*}$\orcidA{},  Luis Ramirez Camargo $^{2}$\orcidB{}, Rüdiger Franke $^{1}$, Ann Nowé $^{3}$\orcidC{}, Maarten Messagie $^{2}$\orcidD{}}

\AuthorNames{Glenn Ceusters, Luis Ramirez Camargo, Rüdiger Franke, Ann Nowé, Maarten Messagie}

\address{%
$^{1}$ \quad ABB, Hoge Wei 27, 1930 Zaventem, Belgium; glenn.ceusters@be.abb.com; ruediger.franke@de.abb.com;\\
$^{2}$ \quad Vrije Universiteit Brussel (VUB), ETEC-MOBI, Pleinlaan 2, 1050 Brussels, Belgium; glenn.leo.ceusters@vub.be; Luis.Ramirez.Camargo@vub.be; maarten.messagie@vub.be;\\
$^{3}$ \quad Vrije Universiteit Brussel (VUB), AI-lab, Pleinlaan 2, 1050 Brussels, Belgium; gceusters@ai.vub.ac.be; ann.nowe@ai.vub.ac.be; \\
}

\corres{\textbf{Correspondence:} glenn.ceusters@be.abb.com}

\abstract{Reinforcement learning (RL) is a promising optimal control technique for multi-energy management systems. It does not require a model \textit{a priori} - reducing the upfront and ongoing project-specific engineering effort and is capable of learning better representations of the underlying system dynamics. However, \textit{vanilla} RL does not provide constraint satisfaction guarantees - resulting in various potentially unsafe interactions within its safety-critical environment. In this paper, we present two novel safe RL methods, namely SafeFallback and GiveSafe, where the safety constraint formulation is decoupled from the RL formulation. These provide hard-constraint, rather than soft- and chance-constraint, satisfaction guarantees both during training a (near) optimal policy (which involves exploratory and exploitative, i.e. greedy, steps) as well as during deployment of any policy (e.g. random agents or offline trained RL agents). This without the need of solving a mathematical program, resulting in less computational power requirements and a more flexible constraint function formulation (no derivative information is required). In a simulated multi-energy systems case study we have shown that both methods start with a significantly higher utility (i.e. useful policy) compared to a \textit{vanilla} RL benchmark and Optlayer benchmark (94,6\% and 82,8\% compared to 35,5\% and 77,8\%) and that the proposed SafeFallback method even can outperform the \textit{vanilla} RL benchmark (102,9\% to 100\%). We conclude that both methods are viably safety constraint handling techniques applicable beyond RL, as demonstrated with random policies while still providing hard-constraint guarantees.}


\usepackage{graphicx}
\usepackage{pdflscape}
\usepackage{longtable}
\usepackage{cleveref}
\usepackage{wrapfig}
\usepackage[ruled,vlined]{algorithm2e}
\usepackage[official]{eurosym}
\usepackage{makecell}
\usepackage{adjustbox}
\begin{document}

\section*{Highlights} 
\begin{itemize}
    \item A (near-to) optimal multi-energy management policy can be learned safely
    \item Any reinforcement learning algorithm can be used safely
    \item Constraint functions increase the initial utility of the policy
    \item Constraints can be formulated independently from the (optimal) control technique
    \item Better policies can be found starting with an initial safe fallback policy
\end{itemize}


\section{Introduction}
\par Energy systems continue to become increasingly interconnected with each other as the energy technologies that allow for this sector coupling are more mature and are being more widely implemented. This allows for an integrated control strategy that further can enhance the overall efficiency and performance of these so-called multi-energy, -carrier, -commodity or -utility systems. Furthermore, these multi-energy systems allow for the utilization of the flexibility (i.e. storage, controllable loads) within and across all energy carriers. This integrated control strategy then typically \cite{Fabrizio2009Trade-offSystems} has an economic or environmental oriented objective function or a combination thereof and therefor being multi-objective. 

\par To ensure the optimum or Pareto optimum level of operation of such multi-energy systems, specific set-points are required to establish and maintain the desired objective (e.g. minimization of the energy costs or $CO_{2}$-equivalent emissions) while still fulfilling all system constraints \cite{Engell2007FeedbackOperation}. As flexibility utilization exhibits dynamic behaviour and introduces a dependency between successive time steps, optimisation across (or considering) numerous time steps is necessary. Additionally, multiple uncertainties (i.e. variation in demands, pricing and weather) need to be managed so that the stability of the multi-energy system is preserved.

\par
\begin{figure}[H]
\centering
\begin{subfigure}{.5\textwidth}
  \centering
  \includegraphics[width=1\linewidth]{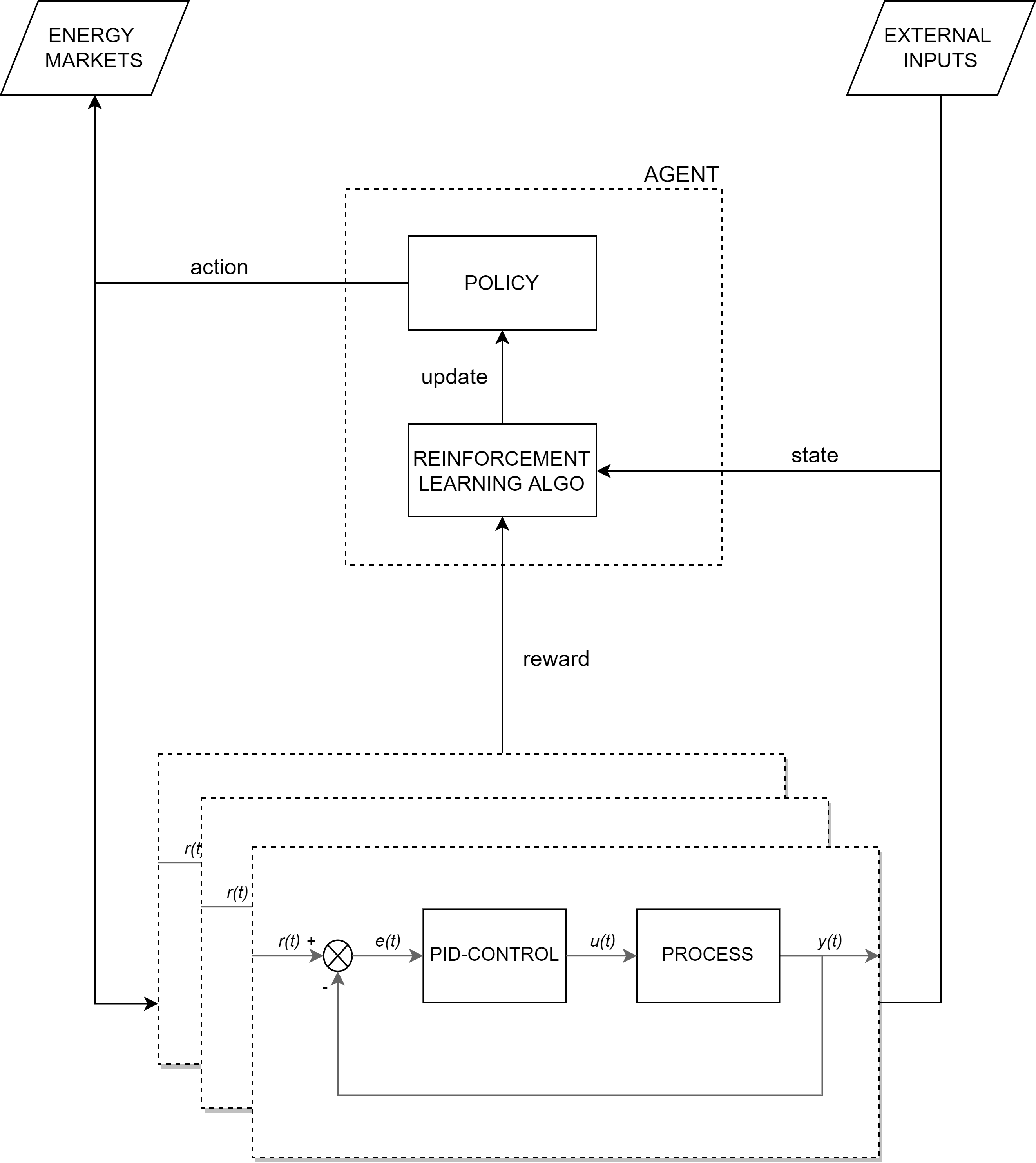}
  \caption{\textit{vanilla} (unsafe) reinforcement learning}
  \label{fig:unsaferl}
\end{subfigure}%
\begin{subfigure}{.5\textwidth}
  \centering
  \includegraphics[width=1\linewidth]{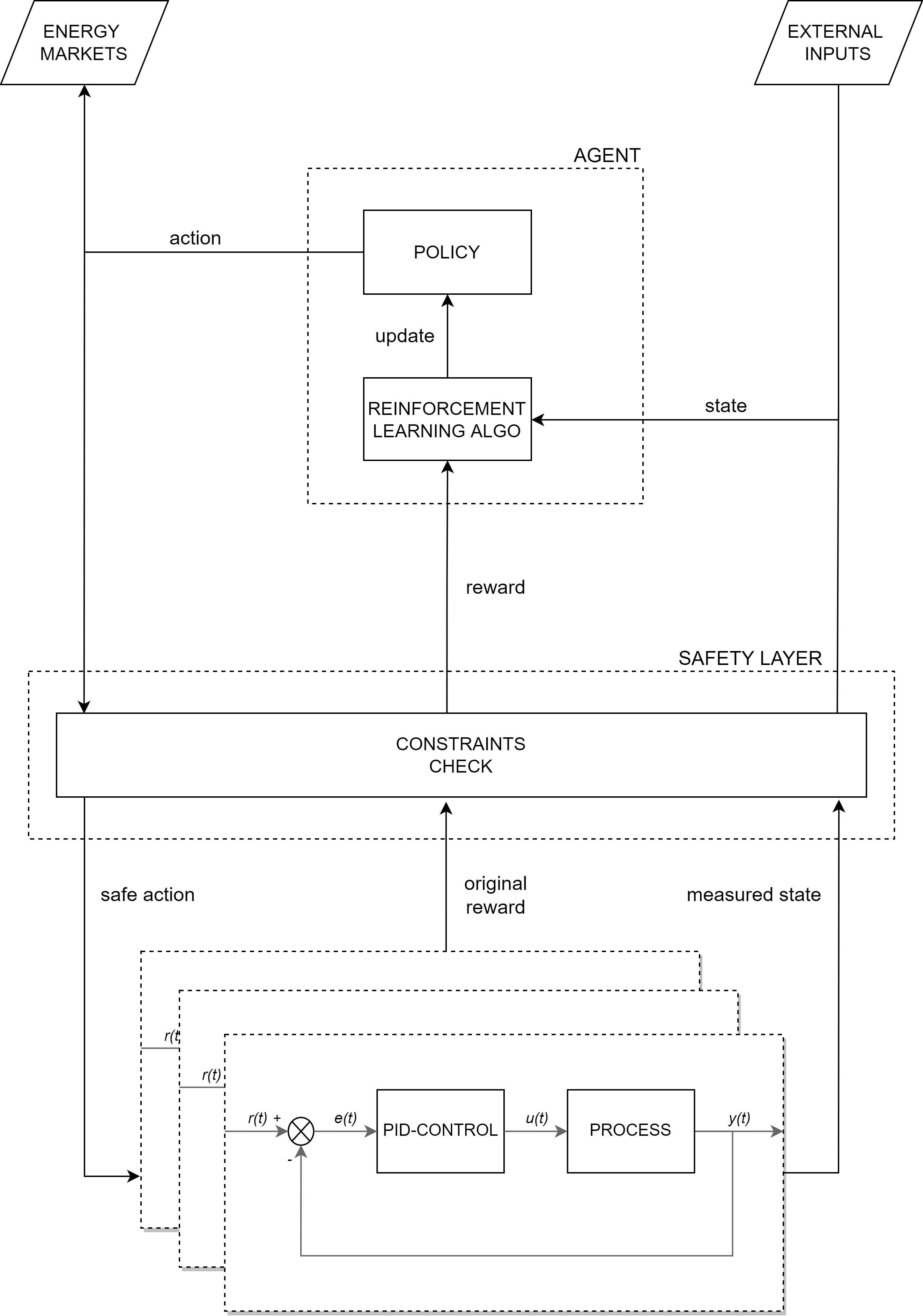}
  \caption{shielded (safe) reinforcement learning}
  \label{fig:saferl}
\end{subfigure}
\vspace*{10pt}
\caption{block diagrams comparison: the feasibility of the RL agent's actions, being in a given state, are always checked against the \textit{a priori} constraint functions acting as a safety layer - shielding the environment from unsafe (control) actions. Note that, we assume that the continuous unconstrained error handling (i.e. minimization of the difference between a desired set-point and a measured process variable) is performed by proportional-integral-derivative (PID) controllers.}
\label{fig:blockdiagrams}
\end{figure}

 Model-predictive control (MPC) and  reinforcement learning (RL) have recently been benchmarked within such a multi-energy management system context \cite{Ceusters2021Model-predictiveStudies}. \citeauthor{Ceusters2021Model-predictiveStudies} showed that RL-based energy management systems do not require a model \textit{a priori} and that they can outperform linear MPC-based energy management systems after training. However, \textit{vanilla} RL (see \autoref{fig:unsaferl}) performs a large number of potentially unsafe interactions within its environment, which is unacceptable in many real-world applications. For example, in a multi-energy system, neglecting the crucial energy balance constraint could result in either under- or over-production.
For most power systems this imbalance could result in exceeding the maximum power capacity to or from the grid - especially relevant with the expected large-scale integration of electric vehicles. Moreover, this imbalance is particularly problematic for energy systems that are not connected to a larger power distribution grid or a district heating system and therefore lack a higher level of control. In this case the imbalance could lead to loss of user comfort (e.g. power or heat outages).

\par Therefore, our goals is to ensure that \textit{every} interaction with the underlying environment (a multi-energy system in our case study) satisfies a given set of safety constraints, \textit{independently} of the used (optimal) control technique (see \autoref{fig:saferl}). This compared to formulating a specific safe RL algorithm which allows that future - presumably better - optimization algorithms can easily be used instead.

\subsection{Contribution and outline}
\par Our contributions can, to the best of our knowledge, be listed as the following:
\begin{itemize}
    \item Online model-free safe RL method which provides hard-constraint, rather than soft- and chance-constraint, satisfaction guarantees that has a significantly higher initial utility;
    \item Decoupling architecture of safety constraint formulations from the RL formulation so that future - presumably better - optimization algorithms can easily be used instead;
    \item Hard-constraint satisfaction without the need of solving a mathematical program, resulting in less computational power requirements and a more flexible constraint function formulation (no derivative information is required);
    \item Demonstration of safe RL-based energy management on a detailed multi-energy system simulation environment.
\end{itemize}

\par In \autoref{chapter2} related work is discussed and our research question is formulated, \autoref{chapter3} introduces the proposed methodologies, while in \autoref{chapter4} the tool chain, the simulated multi-energy system environment, the safety layer, the RL agent and the evaluation procedure are presented. Furthermore, \autoref{chapter5} discusses the results and provides directions for future work while \autoref{chapter6} presents the conclusion. Finally, \hyperref[Appendix A]{Appendix~\ref*{Appendix A}}  shows time series visualizations of the different policies, \hyperref[Appendix B]{Appendix~\ref*{Appendix B}} the pseudo-code of the specific RL agent (TD3) and \hyperref[Appendix C]{Appendix~\ref*{Appendix C}} the run-time statistics.

\section{Related work} \label{chapter2}
\par In recent years, there have been numerous of works that proposed RL for various applications within energy systems as reviewed by e.g.  \cite{Cao2020ReinforcementReview}, \cite{Yang2020ReinforcementSurvey} and \cite{Perera2021ApplicationsSystems}. The majority of these applications can be classified under a broader energy management problem. RL based energy management systems have even been proposed and demonstrated within the more specific (and arguably more challenging) multi-energy systems context. For example, \citeauthor{Rayati2015ApplyingGrid} \cite{Rayati2015ApplyingGrid} were some of the first to apply RL, specifically Q-learning \cite{watkins1989learning}, for the energy management of a simulated multi-energy residential building, which they later extended with demand-side management capabilities \cite{Sheikhi2016DemandSystems}. Posteriorly, \citeauthor{Mbuwir2018BatteryLearning} \cite{Mbuwir2018BatteryLearning} successfully applied RL (Q-learning) for a battery energy management system within a simulated residential multi-energy system. They inlcuded a back-up policy to over-rule the actions of the RL agent in case of constraint violation. Furthermore, \citeauthor{Wang2019Bi-levelSystem} \cite{Wang2019Bi-levelSystem} used a path tracking interior point method and a RL algorithm (Q-learning) for a bi-level interactive decision-making model with multiple agents in a regional multi-energy system. A multi-agent RL (Q-learning) approach was also proposed by \citeauthor{Ahrarinouri2020Multi-AgentBuildings} \cite{Ahrarinouri2020Multi-AgentBuildings} and this for the energy management of a simulated residential multi-energy system. Around the same time, \citeauthor{Ye2020Model-FreeLearning} \cite{Ye2020Model-FreeLearning} proposed the usage of a deep RL algorithm, specifically a deep deterministic policy gradient (DDPG) \cite{Lillicrap2015ContinuousLearning} with a prioritized experience replay strategy, again within a simulated residential multi-energy system. 
\par Moreover, \citeauthor{Xu2021Multi-energyEvolution} \cite{Xu2021Multi-energyEvolution} demonstrated an industrial multi-energy scheduling framework using a RL (Q-learning) based differential evolution approach that adaptively determines the optimal mutation strategy and its associated parameters. While \citeauthor{Zhu2022EnergyPark} \cite{Zhu2022EnergyPark} demonstrated a multi-agent deep RL energy management system, using multi-agent counterfactual soft actor-critic (mCSAC) \cite{Pu2021DecomposedLearning}, for a simulated multi-energy industrial park. Furthermore, \citeauthor{Ceusters2021Model-predictiveStudies} \cite{Ceusters2021Model-predictiveStudies} presented an on- and off-policy multi-objective model-free RL approach, using proximal policy optimisation (PPO \cite{Schulman2017ProximalAlgorithms}) and twin delayed deep deterministic policy gradient (TD3 \cite{Fujimoto2018AddressingMethods}) and they did benchmark this against a linear MPC - both derived from the general optimal control problem. They showed, on two separate simulated multi-energy systems, that the RL agents offer the potential to match and outperform the MPC. While, \citeauthor{Zhang2021SoftEnergy} presented a series of works \cite{Zhang2019DeepEnergy, Zhang2020DynamicApproach, Zhang2021SoftEnergy} for the (near-to) optimal scheduling of an integrated energy system (a.k.a. multi-energy system) using deep reinforcement learning both for a single- and multi-objective and \citeauthor{Zhang2022AFreshwater} \cite{Zhang2022AFreshwater} lather extended this for distributed multi-energy systems using multi-agent deep reinforcement learning.
\par However, as RL inherently requires the interaction with its environment, adequate measures are required to avoid the violation of the environmental specific constraints both during online training as during pure policy execution (e.g. after training a policy offline). All the works above have, knowingly (and thus reported as such) or non-knowingly, either neglected these specific environmental constraints or greatly simplified them - limiting the real-world use cases.
\par In one of the first attempts to combine both hard-constraint satisfaction and RL in energy systems, \citeauthor{Venayagamoorthy2016DynamicMicrogrid} \cite{Venayagamoorthy2016DynamicMicrogrid} presented an intelligent dynamic energy management system for a smart microgrid using an action-dependent heuristic dynamic program, a type of adaptive critic design-based controller. They furthermore used an evolutionary algorithm to improve the dispatch solution over time and rejected candidate solutions that did not satisfy the critical load fulfillment constraint relying on power balancing rules and an initial decision-tree energy management system.
Furthermore, \citeauthor{Zhang2020AInformation} \cite{Zhang2020AInformation} proposed a bi-level power management system of networked microgrids in an electric distribution systems. At the first level, a cooperative agent employs an adaptive model-free RL algorithm, to find the optimal retail price signals for the microgrids. While on the second level, each model-based microgrid controller solves a constrained mixed integer nonlinear program, based on the received price signal from the RL agent. 
Also, \citeauthor{Zhao2020CooperativeLearning} \cite{Zhao2020CooperativeLearning} proposes a knowledge-assisted RL framework by combining a low-fidelity analytical model with a RL agent for a cooperative wind farm control problem. When the RL agent selects a naive action, a constraint action is calculated by solving an optimization problem using that analytical model.
However, in all these cases it remains a heavy reliance on \textit{a prior} physical models that are used in a separate optimization problem. Such developments contain a presumed transition function, a separated objective function (separated from the reward signal, which can introduce bias) and constraint functions - and thus not \textit{only} constraint functions.
\par Nevertheless, recently and independent from this work, \citeauthor{Park2022DIP-QL:Systems} \cite{Park2022DIP-QL:Systems} devised a novel RL algorithm, inspired by \texttt{OptLayer} \cite{Pham2018OptLayerWorld}, namely distance-based incentive/penalty Q-learning (DIP-QL) which also does not assume an \textit{a prior} transition function and only uses constraint functions to provide hard-constraint guarantees as they demonstrated on a microgrid control problem. Yet, it uses a deep Q-learning algorithm as the backbone for their proposed method - where we propose to decouple the constraint handling from the RL algorithm so that future - presumably better - optimization algorithms (as our framework is not limited to RL) can easily be used instead.

\par Concerning safe RL beyond the energy systems management field, \citeauthor{Garcia2015ALearning} \cite{Garcia2015ALearning}, in a comprehensive review, identified and classified two broader approaches: (1) modifying the optimality criteria with a safety factor; (2) modifying the exploration process by incorporating external knowledge or the guidance by a risk metric. More recently, \citeauthor{Dulac-Arnold2021ChallengesAnalysis} \cite{Dulac-Arnold2021ChallengesAnalysis} identified nine challenges that must be addressed to implement RL in real-world problems, including safety constraint violation. Furthermore, \citeauthor{Brunke2021SafeLearning} \cite{Brunke2021SafeLearning} provided a broader safe learning review across both the control theory research space as well as the RL research space. More specifically, they showed (1) learning-based control approaches that start with an \textit{a priori} model to safely improve the policy under the uncertain system dynamics, (2) safe RL approaches that do not need a model or even constraints in advance - but then also do not provide strict safety guarantees (yet encourages safety or robustness), and (3) approaches that provide safety certificates of any learned control policy.

\par The reviewed literature shows that RL is a promising and widely proposed approach for various applications in energy systems (and other domains, not discussed here), as well as for energy management systems specifically. It is also clear that the transition of RL towards real-world applications is not trivial and requires special attention concerning safety. Multiple safe (reinforcement) learning approaches exist, ranging in level of safety namely (from lower to higher level): (1) soft-constraint satisfaction, (2) chance-constraint satisfaction and (3) hard-constraint satisfaction. However, model-free safe RL where the safety constraint formulation is decoupled from the RL formulation and which provides hard-constraint, rather than soft- and chance-constraint, satisfaction guarantees both during training a (near) optimal policy (which involves exploratory and exploitative, i.e. greedy, steps) as well as during deployment of any policy (e.g. random agents or offline trained RL agents). This without the need of solving a mathematical program has - to the best of our knowledge - never been proposed and demonstrated for the energy management of multi-energy systems.

\section{Proposed methodology} \label{chapter3}
\par Following the standard RL formulation of the state-value function, yet extending this towards constraints subjection, the objective is to find a policy \( \pi \), which is a mapping of states, \( s \in S\), to actions, \(a \in A(s)\), that maximizes an expected sum of discounted rewards and is subject to constraint sets \( X\) and \( U\):
\begin{subequations}
\begin{align}
    \label{equation1a}
    \max_\pi &\Bigg(E_\pi\bigg\{ \sum_{k = 0}^{\infty} \gamma^k R_{t+k+1} \bigg\}\Bigg)\\[1em]
    \label{equation1b}
    s.t. \quad & s_t = s && t \in \mathbb{T}_{0}^{+\infty} = 0,\dots,+\infty\\
    \label{equation1c}
    \quad & s_t \in X && t \in \mathbb{T}_{0}^{+\infty} = 0,\dots,+\infty\\
    \label{equation1d}
    \quad & a_t \in U && t \in \mathbb{T}_{0}^{+\infty} = 0,\dots,+\infty
\end{align}
\end{subequations}
where \(E_\pi\) is the expected value, following the policy \(\pi\), of the rewards \(R\) discounted with the discount factor \(\gamma\) over an infinite sum at any time step \(t\).
\par Note that \autoref{equation1a} is a discrete\footnote{as the continuous error handling is performed by PID-controllers, see \autoref{fig:blockdiagrams}} time-invariant infinite-horizon stochastic optimal control problem, as also indicated in \cite{Ceusters2021Model-predictiveStudies}, yet differs from the standard formulation of RL due to the \textit{a priori} constraint functions in the sets \( X\) and \( U\). We acknowledge that methods without \textit{a priori} constraint functions exists, yet this can - under the state-of-the-art - only reach safety level 2 at best (chance-constraint satisfaction) \cite{Brunke2021SafeLearning}. Rather than proposing a specific safe RL algorithm, we propose to decouple the \textit{a priori} constraint function formulation from the (RL) agent so that any (new RL) algorithm can be used - while always guarantying the hard-constraint satisfaction. Although the proposed algorithms are conceptually simple, we will later show its effectiveness on a multi-energy system (which includes a non-grid connected thermal system).

\subsection{SafeFallback method} \label{chapter 3.1}
\par The first method we propose relies on an \textit{a priori} safe fallback policy \(\pi^{safe}\), which typically can be derived through classic control theory in the form of a set of hard-coded rules such as a simple rule-based policy (e.g. a priority-based energy management strategy - which is commonly available or easily constructible - see \autoref{chapter 4.4} for the safe fallback policy of the considered case study). As we furthermore assume that the constraint functions are given, we can simply check if the selected actions \(a\) while in state \(s\) satisfy the constraint conditions. When the constraints conditions are satisfied, the selected actions \(a\) are \textit{considered} to be safe actions \( a^{safe} \) and are then executed in the environment so that the next state \( s' \), the reward \( r \) and done signal \(d\) are observed - which is the regular experience tuple \( (s,a^{safe},r,s',d) \) for the RL agent. However, if the constraint conditions are violated, the selected action \(a\) is overruled by the safe action \( a^{safe} \) using the \textit{a priori} safe fallback policy \(\pi^{safe}\). Now not only the experience tuple \( (s,a^{safe},r,s',d) \) is formed, but also the experience tuple \( (s,a,r-c,s',d) \) containing the infeasible action and additional negative reward (i.e. cost, \(c\)). The pseudo-code is given in \autoref{algo: SafeFallback}. \vspace{10pt}

\begin{algorithm}[H]
\DontPrintSemicolon
\SetAlgoLined
 \nl Input: initialize RL algorithm, initialize constraint functions in sets \(X\) and \(U\), initialize safe fallback policy \(\pi^{safe}\) \;
 \nl \For{\( k=0,1,2,\dots \)}{
 \nl Observe state \(s\) and select action \(a\)\;
 \nl \eIf{constraint check = True}{keep selected action \(a\) as safe action \(a^{safe}\)}{get safe action \(a^{safe}\) from safe fallback policy \(\pi^{safe}\) \;} 
 \nl Execute \( a^{safe} \) in the environment \;
 \nl Observe next state \( s' \), reward \( r \) and done signal \( d \) to indicate whether \( s' \) is terminal \;
 \nl Give experience tuple \( (s,a^{safe},r,s',d) \) \textbf{and if} \(a^{safe} \neq a:\) \( (s,a,r-c,s',d) \) with cost \(c\) \;
 \nl If \( s' \) is terminal, reset environment state \;
 }
 \caption{SafeFallback}\label{algo: SafeFallback}
\end{algorithm}

\subsection{GiveSafe method}
\par
\begin{figure}[H]
    \centering
    \includegraphics[width=\textwidth]{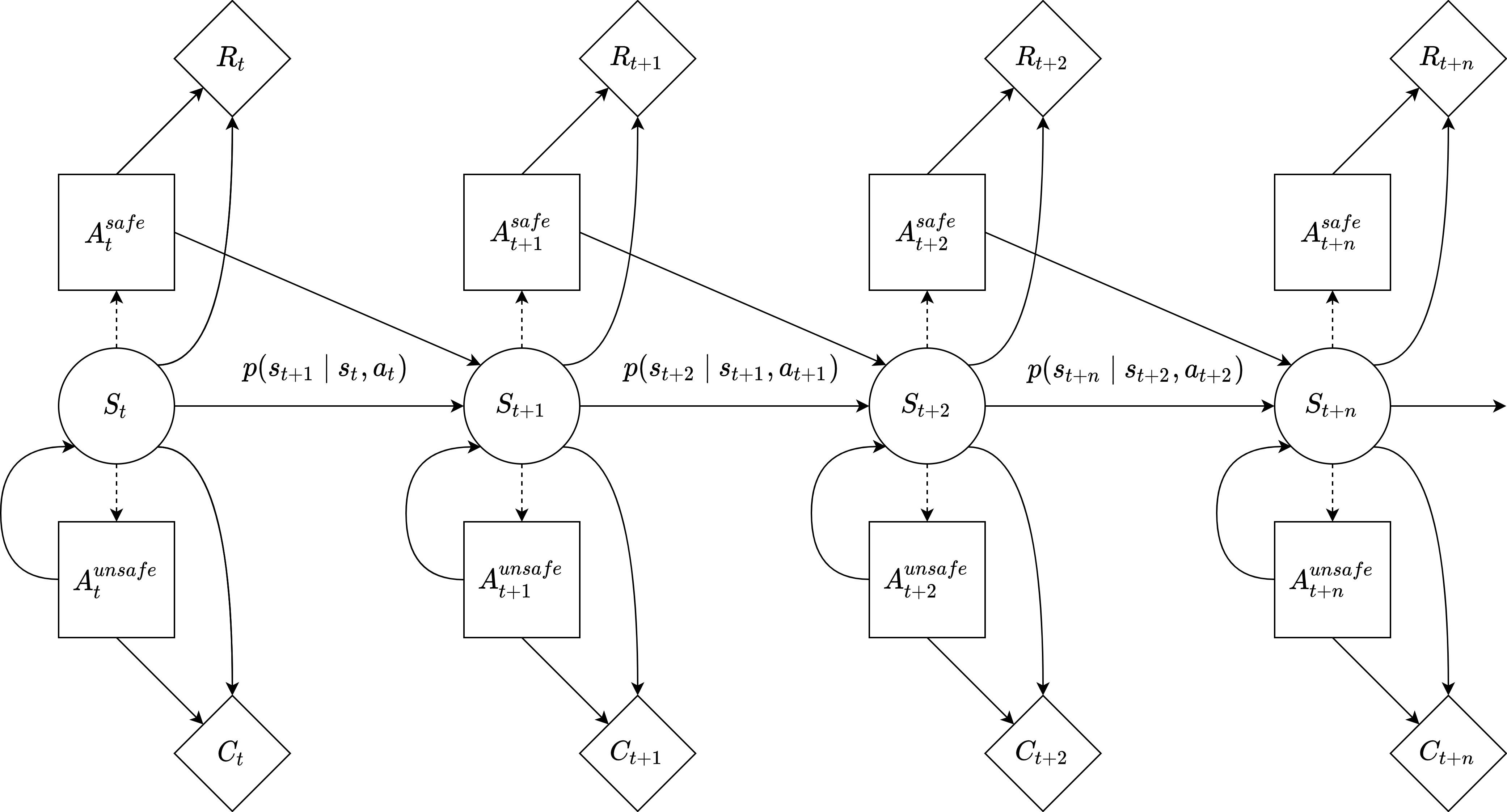}
    \caption{Markov Chain of \autoref{algo: GiveSafe}. When the selected action is infeasible (does not satisfy to the constraints), that unsafe action \(A^{unsafe}_t\) is not executed in the environment so no transition to the next state \(S_{t+1}\) is observed and a cost \(C_t\) is given. When the selected action is feasible (satisfies the constraints), that safe actions \(A^{safe}_t\) is executed in the environment so a transition to the next state \(S_{t+1}\) is observed with probability \(p(s_{t+1} \mid s_t,a_t)\) and  a reward \(R_t\) is given.}
    \label{fig: markov chain givesafe}
\end{figure}

 Our second method does not require an \textit{a priori} safe fallback policy, yet relies on the RL agent itself to pass safe actions \( a^{safe} \) - which can again be checked by the given constraint conditions. If the selected actions \(a\) while in state \(s\) passes the constraint check, the safe actions \( a^{safe} \) get executed in the environment and a regular experience tuple \( (s,a^{safe},r,s',d) \) is received. However, if the constraints get violated the RL agent receives the experience tuple \( (s,a,r-c,s,d) \). Hence, the transition towards the next state is not observed (as the infeasible action is not executed) and a cost \(c\) (i.e. negative reward) is given. The RL agent then selects a new action \(a\) and a new constraint check is done. This is repeated until the constraint check gets passed and the selected action is considered to be a safe action \( a^{safe} \). This safe action is then executed in the environment and a regular experience tuple is received. The pseudo-code is given in \autoref{algo: GiveSafe} and a graphical representation, in the form of a Markov Chain, in \autoref{fig: markov chain givesafe}. \vspace{10pt}

\begin{algorithm}[H]
\DontPrintSemicolon
\SetKwRepeat{Do}{do}{while}%
\SetAlgoLined
 \nl Input: initialize RL algorithm, initialize constraint functions in sets \(X\) and \(U\)\;
 \nl \For{\( k=0,1,2,\dots \)}{
 \nl Observe state \(s\) and select action \(a\)\;
 \nl \eIf{constraint check = True}{keep selected action \(a\) as safe action \(a^{safe}\)}{
 \nl \While{constraint check = False}{give experience tuple \( (s,a,c,s,d) \) with cost \(c\) \;
 agent selects new action \(a\)\;
 check constraints} \Return safe action \(a^{safe}\)}
 \nl Execute \( a^{safe} \) in the environment \;
 \nl Observe next state \( s' \), reward \( r \) and done signal \( d \) to indicate whether \( s' \) is terminal \;
 \nl Give experience tuple \( (s,a^{safe},r,s',d) \) \;
 \nl If \( s' \) is terminal, reset environment state \;
 }
 \caption{GiveSafe}\label{algo: GiveSafe}
\end{algorithm}

\section{Simulated case study} \label{chapter4}
\subsection{Toolchain}
\par A multi-energy systems simulation model, that was developed in \cite{Ceusters2021Model-predictiveStudies}, was used as it allowed for the verification of the safety critical operation (e.g. if all energy demands are fulfilled) of the multi-energy system without consequences (i.e. without the risk of violating real-life constraints and its associated harm). It is a \texttt{Modelica} \cite{Mattsson1998PhysicalModelica} model, as it allowed for the convenient construction of the real-life \textit{presumed} system dynamics using multi-physical first-principle equations and due to the available highly specialized libraries, elementary components and its object-oriented nature. 

\par
\begin{figure}[ht]
    \centering
    \includegraphics[scale=0.125]{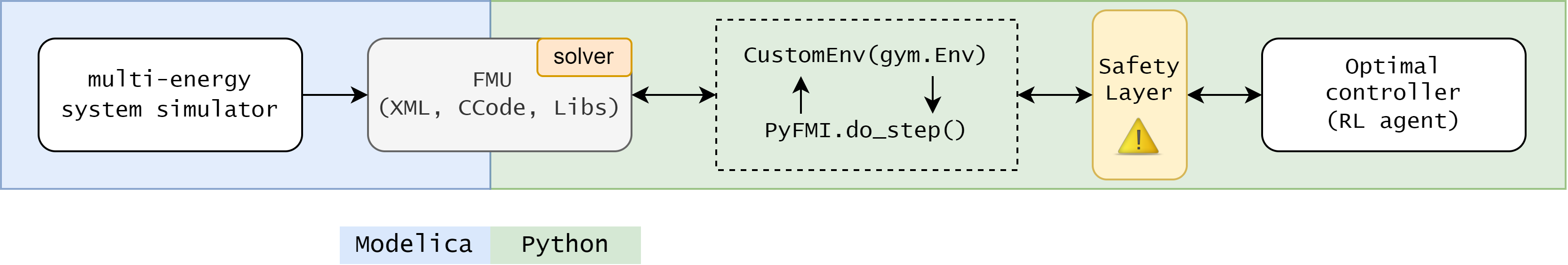}
    \caption{architecture of the tool-chain}
    \label{fig: architecture}
\end{figure}

 This \texttt{Modelica} model is then exported as a co-simulation \textit{functional mock-up unit} (FMU), similar to \cite{Graber2017FromProblems} , and wrapped into an \texttt{OpenAI} gym \textit{environment} \cite{Brockman2016OpenAIGym} in \texttt{Python}, similar to \cite{Lukianykhin2019ModelicaGym:Models, Ceusters2021Model-predictiveStudies}. The architecture of the tool-chain is shown in \autoref{fig: architecture}. Notice that the Differential Algebraic Equations solver is part of the co-simulation FMU and that the \texttt{do\_step()} method in \texttt{PyFMI} \cite{Andersson2016PyFMI:Interface} is used over \texttt{simulate()} - again as in \cite{Ceusters2021Model-predictiveStudies} due to the significant run-time speed-up when initialized properly.

\subsection{Simulation model}
\par The considered multi-energy system is similar to the one  from \cite{Ceusters2021Model-predictiveStudies}, yet without the gas turbine (back-up genset)\footnote{as it is not required, nor does it provide additional value, to test the proposed methods}, and has the following structure:

\begin{figure}[!hb]
	\centering
	\includegraphics[scale=0.1]{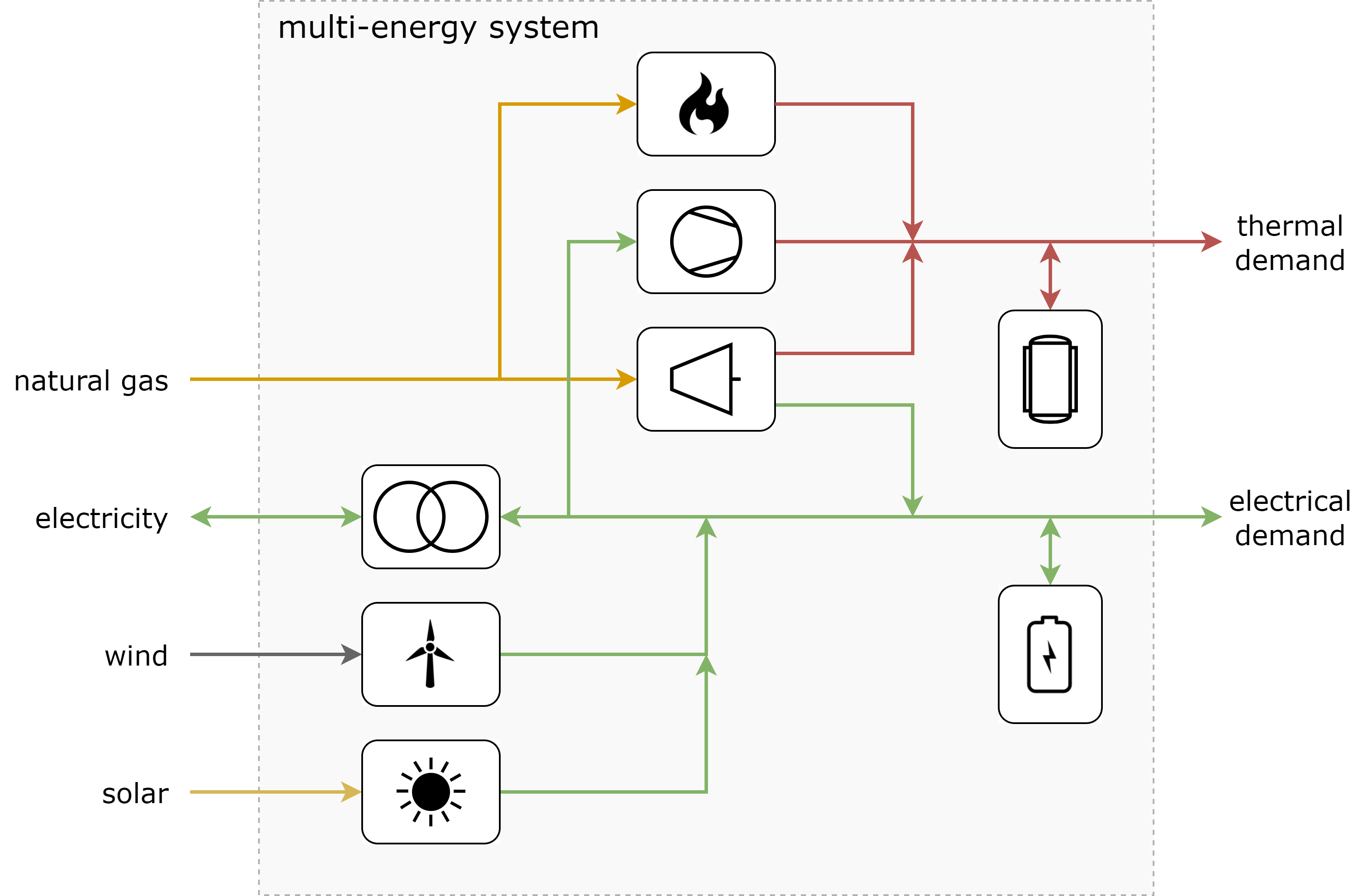}
	\caption{structure of the simulated multi-energy system}
	\label{fig: case I structure}
\end{figure}

\par It includes (from left to right, from top to bottom): an electric transformer, a wind turbine, a photovoltaic (PV) installation, a natural gas boiler, a heat pump (HP), a combined heat and power (CHP) unit, a thermal energy storage system (TESS) and a battery energy storage system (BESS). The dimensions of the considered multi-energy system are also from \cite{Ceusters2021Model-predictiveStudies} and are summarised in \autoref{tab: mes dimensions}.
\begin{table}[!ht]
    \centering
    \begin{tabular}{c|c|c|c|c|c}
    \rowcolor[HTML]{efefef} 
    \textbf{Energy asset} & \textbf{Input} & \textbf{Output} & \textbf{P\textsubscript{nom}} & \textbf{P\textsubscript{min}} & \textbf{E\textsubscript{nom}} \\
    \hline
    transformer & elec & elec & \(+\infty\) & \(-\infty\) & \\
    wind turbine & wind & elec & 0.8 MW\textsubscript{e} & 1.5 \% & \\
    solar PV & solar & elec & 1.0 MW\textsubscript{e} & 0 \% & \\
    boiler & CH\textsubscript{4} & heat & 2.0 MW\textsubscript{th} & 10 \% & \\
    heat pump & elec & heat & 1.0 MW\textsubscript{th} & 25 \% & \\
    CHP & CH\textsubscript{4} & heat & 1.0 MW\textsubscript{th} & 50 \% & \\
     & CH\textsubscript{4} & elec & 0.8 MW\textsubscript{e} & 50 \% & \\
    TESS & heat & heat & +0.5 MW\textsubscript{th} & -0.5 MW\textsubscript{th} & 3.5 MWh \\
    BESS & elec & elec & +0.5 MW\textsubscript{e} & -0.5 MW\textsubscript{e} & 2.0 MWh 
    \end{tabular}
    \caption{dimensions of the multi-energy system}
    \label{tab: mes dimensions}
\end{table}

\subsection{Safety layer} \label{chapter4.3}
\par The constraint functions, \autoref{equation1c} and \autoref{equation1d}, are in this case study specifically (see note of \autoref{equation3c} regarding modeling effort) :
\begin{subequations}
\begin{align}
    \label{equation2a}
    Q^{min}_{boil} \times \gamma^{t}_{boil} \leq Q^{t}_{boil} \leq Q^{max}_{boil} \times \gamma^{t}_{boil} &&\forall t&, \hspace{5pt} \gamma^{t}_{boil} \in \{0,1\} \\
    \label{equation2b}
    \begin{bmatrix}
    Q^{min}_{hp} \\
    P^{min}_{hp}
    \end{bmatrix}  \times \gamma^{t}_{hp} \leq
    \begin{bmatrix}
    Q^{t}_{hp} \\
    P^{t}_{hp}
    \end{bmatrix} \leq 
    \begin{bmatrix}
    Q^{max}_{hp} \\
    P^{max}_{hp}
    \end{bmatrix}  \times \gamma^{t}_{hp} &&\forall t&, \hspace{5pt} \gamma^{t}_{hp} \in \{0,1\} \\
    \label{equation2c}
    \begin{bmatrix}
    Q^{min}_{chp} \\
    P^{min}_{chp}
    \end{bmatrix}  \times \gamma^{t}_{chp} \leq
    \begin{bmatrix}
    Q^{t}_{chp} \\
    P^{t}_{chp}
    \end{bmatrix} \leq 
    \begin{bmatrix}
    Q^{max}_{chp} \\
    P^{max}_{chp}
    \end{bmatrix}  \times \gamma^{t}_{chp} &&\forall t&, \hspace{5pt} \gamma^{t}_{chp} \in \{0,1\} \\
    \label{equation2d}
    Q^{min}_{tess} \leq Q^{t}_{tess} \leq Q^{max}_{tess}  &&\forall t& \\
    \label{equation2e}
    P^{min}_{bess} \leq P^{t}_{bess} \leq P^{max}_{bess}  &&\forall t& \\
    \label{equation2f}
    Q^{t}_{production} = Q^{t}_{demand} &&\forall t&
\end{align}
\end{subequations}
where \(Q_{boil}^t\), \(Q_{hp}^t\), \(Q_{chp}^t\) and \(Q_{tess}^t\) are the thermal powers of the natural gas boiler, the heat pump, the combined heat and power unit (CHP) and the thermal energy storage system (TESS) respectively while \(P_{hp}^t\), \(P_{chp}^t\) and \(P_{bess}^t\) represent the electrical powers of the heat pump, CHP and battery energy storage system (BESS), all constraint by its associated minimal and maximal power (see \autoref{tab: mes dimensions}). Furthermore, \(\gamma^{t}_{boil}\), \(\gamma^{t}_{hp}\) and \(\gamma^{t}_{chp}\) are binary variables that turn on/off the given asset (i.e. as the minimal powers are not zero). Yet these constraints, i.e. \autoref{equation2a} till \autoref{equation2e}, are easily handled by the dimensions of the (control) action space itself, i.e. \autoref{equation4b} till \autoref{equation4f}, and therefor do not require a specific constraint check. 
\par While the electrical energy balance is always fulfilled (given the assumption that the electrical grid connection is sufficiently large), the thermal energy balance (\autoref{equation2f}) does require special attention in order to achieve hard-constraint satisfaction. No additional constraints are considered in this case study (e.g. ramping rates, minimal run- and down-time) as energy balance equations are considered to be the most limiting constraints in energy management problems. Writing out the thermal energy balance in more detail and relaxing the equality constraint formulation (towards an inequality constraint) then becomes:
\begin{subequations}
\begin{align}
    \label{equation3a}
    \left| Q^{t}_{boil} + Q^{t}_{hp} + Q^{t}_{chp} + Q^{t}_{tess} - Q^{t}_{demand} \right| \leq Q_{tol} && \forall t \\
    | A^{t}_{boil} \cdot \eta_{boil} \cdot Q^{max}_{boil} + A^{t}_{hp} \cdot \frac{COP_{hp}}{COP^{max}_{hp}} \cdot Q^{max}_{hp} + A^{t}_{chp} \cdot \eta^{th}_{chp} \cdot Q^{max}_{chp} \notag\\ 
    \label{equation3b}
    + A^{t}_{tess} \cdot f(SOC^{t}_{tess}) - Q^{t}_{demand} | \leq Q_{tol} && \forall t \\
    | A^{t}_{boil} \cdot f(T^t_{boil}) \cdot Q^{max}_{boil} + A^{t}_{hp} \cdot \frac{f(T^t_{evap}, T^t_{cond})}{COP^{max}_{hp}} \cdot Q^{max}_{hp} \notag\\
    \label{equation3c}
    + A^{t}_{chp} \cdot f(P^t_{chp},Q^t_{chp},T^t_{env}) \cdot Q^{max}_{chp} 
    + A^{t}_{tess} \cdot f(\overline{T^{t}_{tess}}) - Q^{t}_{demand} | \leq Q_{tol} && \forall t
\end{align}
\end{subequations}
where \(A^{t}\) are the (control) actions, \(\eta\) the energy efficiencies, \(COP\) the coefficient of performance, \(SOC\) the state of charge and \(T\) various specific temperatures (i.e. \(T^t_{boil}\) the return temperature to the boiler, \(T^t_{evap}\) the evaporator temperature of the heat pump, \(T^t_{cond}\) the condenser temperature of the heat pump, \(T^t_{env}\) the environmental air temperature and \(\overline{T^{t}_{tess}}\) the average temperature in the stratified hot water storage tank). We set \(Q_{tol}\) to be 15,0\% of the total \(Q^{t}_{demand}\) in the evaluation period, which we acknowledge to be relatively high - yet is chosen to keep the computational complexity low (\hyperref[Appendix C]{Appendix~\ref*{Appendix C}}). Note that the different functions \(f(\cdot)\) from \autoref{equation3c} are typically not trivial to \textit{model} accurately. Therefor we assume the availability of a historical dataset to supervisory learn (using a Random Forest Regression algorithm) the function between the thermal power and the action directly (i.e. \(Q^t_{asset}=f(A^t_{asset}, \chi^t_{asset})\)), with the possibility to include informative exogenous variables \(\chi^t_{asset}\).

\begin{table}[!ht]
    \centering
    \begin{tabular}{c|c|c|c}
    \rowcolor[HTML]{efefef} 
    \textbf{Energy asset} & \textbf{R2-score} & \textbf{MAE} & \textbf{NMAE} \\
    \hline
    boiler & 99.92\% & 7.2 kW & 0.34\% \\
    heat pump & 99.74\% & 6.1 kW & 0.62\% \\
    CHP & 99.86\% & 4.3 kW & 0.38\% \\
    TESS  & 96.22\% & 12.9 kW & 1.37\% \\
    BESS & 99.43\% & 4.6 kW  & 0.46\% 
    \end{tabular}
    \caption{Safety layer model metrics with \texttt{test\_size} of 0.25. Mean Absolute Error (MAE), Normalised Mean Absolute Error (NMAE) by range, i.e. NMAE = MAE / range(actual values)}
    \label{tab: safety layer model metrics}
\end{table}

\subsection{Safe fallback policy} \label{chapter 4.4}
\par As presented in \autoref{chapter 3.1}, \autoref{algo: SafeFallback} relies on an \textit{a prior} safe fallback policy \(\pi^{safe}\) which can be \textit{any} (non-optimal) policy that satisfies the constraints and can typically be provided by domain experts. In our case study this is a simple priority rule: \vspace{5pt}

\begin{center}
\begin{minipage}{0.5\textwidth}
\begin{algorithm}[H]
\DontPrintSemicolon
\SetAlgoLined
 \eIf{\(Q^{t}_{demand} < Q^{min}_{chp}\)}{\(Q^{t}_{chp} = 0\) \\ \(Q^{t}_{boil} = Q^{t}_{demand}\)}{\eIf{\(Q^{t}_{demand} < Q^{max}_{chp}\)}{\(Q^{t}_{chp} = Q^{t}_{demand}\) \\ \(Q^{t}_{boil} = 0\)}{\(Q^{t}_{chp} = Q^{max}_{chp}\) \\ \(Q^{t}_{boil} = Q^{t}_{demand} - Q^{max}_{chp}\)}}
 \caption{safe fallback policy}
\end{algorithm} \vspace{5pt}
\end{minipage}
\end{center}

Note that, for clarity concerns, the policy has been written out in terms of thermal power outputs yet is still converted to actions \(A^{t}_{chp}\) and \(A^{t}_{boil}\) as going from \autoref{equation3a} to \autoref{equation3b}.

\subsection{Energy managing RL agent}
The fully observable discrete-time Markov decision process (MDP) is formulated as the tuple \( \langle S,A,P_a,R_a \rangle \) so that:
\begin{subequations}
\begin{align}
    \label{equation4a}
    S^t = ( E_{th}^t,\ E_{el}^t,\ P_{wind}^t,\ P_{solar}^t,\ X_{el}^t, \ SOC_{tess}^t, \ SOC_{bess}^t, \ h^t, \ d^t  )& & S^t \in S\\[1em]
    \label{equation4b}
    A_{boil}^t = (0,\ A_{boil}^{min} \xrightarrow{} A_{boil}^{max})& & A_{boil}^t \in A \\
    \label{equation4c}
    A_{hp}^t = (0,\ A_{hp}^{min} \xrightarrow{} A_{hp}^{max})& & A_{hp}^t \in A \\
    \label{equation4d}
    A_{chp}^t = (0,\ A_{chp}^{min} \xrightarrow{} A_{chp}^{max})& & A_{chp}^t \in A \\
    \label{equation4e}
    A_{tess}^t = (A_{tess}^{min} \xrightarrow{} A_{tess}^{max})& & A_{tess}^t \in A \\
    \label{equation4f}
    A_{bess}^t = (A_{bess}^{min} \xrightarrow{} A_{bess}^{max})& & A_{bess}^t \in A \\[1em]
    \label{equation4g}
    R_a = - (a \times L_{cost}^t + b \times L_{comfort}^t) - c&
\end{align}
\end{subequations}
where \( E_{th}^t \) is the thermal demand, \( E_{el}^t \) the electrical demand, \( P_{wind}^t \) the electrical wind in-feed, \( P_{solar}^t \) the electrical solar in-feed, \(X_{el}^t\) the electrical price signal (i.e. day-ahead spot price), \(SOC_{tess}^t\) the state-of-charge (SOC) of the TESS, \(SOC_{bess}^t\) the SOC of the BESS, \(h^t\) the hour of the day and \(d^t\) the day of the week all at the \textit{t}-th step, which consitute the state-space \( S \). The action-space \( A \) includes the control set-points from, \( A_{boil}^t \) the natural gas boiler, \( A_{hp}^t \) the heat pump, \( A_{chp}^t \) the CHP unit,  \( A_{tess}^t \) the TESS and \( A_{bess}^t \) the BESS all between a minimum and maximum power rate as shown in \autoref{tab: mes dimensions}.
\par The objective of the energy managing agent is given by the reward function \( R_a \) (i.e. where we try to maximize a negative function, and thus minimize the positive version of that function, in accordance with \autoref{equation1a}) and is the negative loss in energy costs \( L_{cost}^t \) and loss in comfort \( L_{comfort}^t \) both at the \textit{t}-th time-step with multi-objective scaling constants \( a \) and \( b \) and with an additional cost \( c \) when the constraint are \textit{expected} to be violated\footnote{as these unsafe actions are not executed in the environment - see \autoref{algo: SafeFallback} and \autoref{algo: GiveSafe} }. The loss in comfort is defined as \( | E_{th}^t - Q^t | \), where \( Q^t \) is the thermal energy production. The electrical demand and natural gas consumption can always be fulfilled by (buying from) their respective \textit{infinitely} large main grid connection, i.e. within the \texttt{Modelica} simulation model, it is assumed that the grid connections are sufficiently large. Note that the loss in comfort \( L_{comfort}^t \) is bound by the tolerance of the constraints. This term, in the reward function, therefore serves as a fine-tuning mechanism to further minimize the loss in comfort within those bound and to mitigate the modelling error of the constraints itself (see \autoref{tab: safety layer model metrics} for the quality of the constraint functions). 
\par The energy costs \( L_{cost}^t \) is in EUR with scaling factor \( a = 1/10 \), the loss in comfort \( L_{comfort}^t \) is in Watt with scaling factor \( b = 1/5e5 \) and cost \( c = 1 \) in \autoref{algo: SafeFallback} and \( c = -50 + (10 \hspace{5pt} \textbf{if} \hspace{5pt} A_{chp}^t > 0.5 \hspace{5pt} \textbf{else} \hspace{5pt} 0)\) in \autoref{algo: GiveSafe}. The state-space \( S \) is normalized and all actions in the action-space \( A \) are scaled between \( [+1, -1] \).
\par Finally, we use a twin delayed deep deterministic policy gradient (TD3) agent, as it is considered one of the state-of-the-art RL algorithms, from the \texttt{stable baseline} \cite{stable-baselines3} implementations and this with the following hyper-parameters (found after an hyper-parameter optimization study, similar to \cite{Ceusters2021Model-predictiveStudies}, for \autoref{algo: SafeFallback} and \autoref{algo: GiveSafe}). The pseudo-code of the TD3 algorithm is given in \hyperref[Appendix B]{Appendix~\ref*{Appendix B}}.

\begin{table}[!h]
    \centering
    \begin{tabular}{l|c|c|c|c}
    \rowcolor[HTML]{efefef} 
    \textbf{Hyper-parameters: TD3} & \textbf{\autoref{algo: SafeFallback}} & \textbf{\autoref{algo: GiveSafe}} & \textbf{unsafe} & \textbf{optsafe} \\
    \hline
    gamma           & 0.7   & 0.95  &  0.9 & 0.7  \\
    learning\_rate  & 0.000583  & 0.000119  &  0.0003833 & 0.000583  \\
    batch\_size     & 16  & 16  &  100 & 16      \\
    buffer\_size    & 1e6   & 1e5  &  1e5 & 1e6      \\
    train\_freq     & 1e0  & 1e1  &  2e3 & 1e0      \\
    gradient\_steps & 1e0  & 1e1  &  2e3 & 1e0      \\
    noise\_type     & normal    & normal    &  normal & normal   \\
    noise\_std      & 0.183  & 0.791   &  0.329 & 0.183              
    \end{tabular}
    \caption{TD3 hyper-parameters}
\end{table}

\subsection{Evaluation}
\par The performance, in terms of energy cost minimization subject to the (thermal comfort) constraint fulfillment, of the proposed methods \autoref{algo: SafeFallback} and \autoref{algo: GiveSafe} is compared against an unconstrained (and therefore possibly unsafe) RL agent,  the OptLayer RL agent proposed by \citeauthor{Pham2018OptLayerWorld} \cite{Pham2018OptLayerWorld} (identified as related work), as well as  against safe and unsafe random agents - serving as minimal performance benchmarks. We use a year-long training environment, a 15-minutes control horizon (i.e. the energy managing RL agent can select new actions every simulated 15 minutes) and a week-long evaluation environment while participating in a day-ahead electricity market. Any uncertainty (from e.g. demands, prices or wind and solar power generation) is inherently handled by the RL agent, as it is formulated as a discrete time-invariant infinite-horizon \textit{stochastic} optimal control problem (see \cite{Ceusters2021Model-predictiveStudies} for the derivation from a continuous time-varying stochastic system). The model-predictive controller from \cite{Ceusters2021Model-predictiveStudies} is here not considered, as constraints can be formulated directly in the method.

\section{Results and discussion} \label{chapter5}
The simulation results of the objective values (i.e. rewards) are shown, in \autoref{tab: sim results}, both in absolute values as relative to the unconstrained RL benchmark. The \textit{inequality} constraint tolerance, \autoref{equation3c}, is shown relative to the total demand (0\% would mean \textit{equality} constraint satisfaction, i.e. all thermal demand is being fulfilled including any thermal energy storage charging actions).

\begin{table}[!htbp]
    \centering
    \begin{tabular}{l|c|c|c}
    \rowcolor[HTML]{efefef} 
    \textbf{Optimal controller} & \multicolumn{2}{c|}{\textbf{Objective value}} & \textbf{Constraint tolerance}\\
    \hline
    Unsafe TD3 & -5.043 & 100,0\% & 21,0\% \\
    Unsafe Random & -14.223 &  35,5\% & 146,0\% \\
    OptLayer Random & -6.481 & 77,8\% & 15,6\% \\
    OptLayer TD3 & \textbf{-4.850} & \textbf{104,0\%} & 10,4\% \\
    SafeFallback TD3 & -4.899 & 102,9\% & 10,1\% \\ 
    SafeFallback Random & -5.331 & 94,6\% & \textbf{7,0\%} \\
    GiveSafe TD3 & -5.137 & 98,2\% & 10,0\% \\ 
    GiveSafe Random & -6.089 & 82,8\% & 15,0\%
    \end{tabular}
    \caption{5-run average policy performance with a training budget of 15-years worth of time steps per run (i.e. 525.150 time steps per run)}
    \label{tab: sim results}
\end{table}

\par These results (\autoref{tab: sim results}) show that \autoref{algo: SafeFallback} (SafeFallback - 102,9\%) outperforms \autoref{algo: GiveSafe} (GiveSafe - 98,2\%) and the \textit{vanilla} unsafe TD3 benchmark (100\%), yet is slightly worse then OptLayer (104,0\%). This as the \textit{a prior} safe fallback policy has the highest utility before training (94,6\% compared to 82,8\%, 35,5\% and 77,8\%), indicating the additionally given expert knowledge. The additional expert knowledge (of the safe fallback policy of \autoref{algo: SafeFallback} itself) is reflected in the initially higher constraint tolerance as well (7,0\%), yet reaches an acceptable 10,1\% (below the maximum tolerance of 15\%, as set in \autoref{equation3c}). \autoref{algo: GiveSafe} and OptLayer have initially a higher constraint tolerance (15,0\% and 15,6\%), yet reaching approximately the same tolerances (10,0\% and 10,4\%). Both of the proposed methods are therefore, as intended, significantly safer than the \textit{vanilla} TD3 benchmark - which has an initial constraint tolerance of 146,0\% and reaching only 21,0\%. The constraint tolerances converge towards a limit as defined by the multi-objective reward function, \autoref{equation4g}, and its associated scaling factors \(a\) and \(b\) (i.e. energy cost minimization and energy demand fulfillment are conflicting objectives). Note that, OptLayer, initially violates the maximum tolerance of 15\%, as set in \autoref{equation3c}. This happens because OptLayer involves solving a mixed-integer quadratic problem to compute the nearest feasible actions. Moreover, in OptLayer linear analytical approximations are used instead of surrogate functions \(f(\cdot)\) from \autoref{equation3c} with the metrics provided in \autoref{tab: safety layer model metrics}, since there is no derivative information present.

\begin{figure}[H]
    \centering
    \includegraphics[width=0.90\textwidth, height=0.4\textheight]{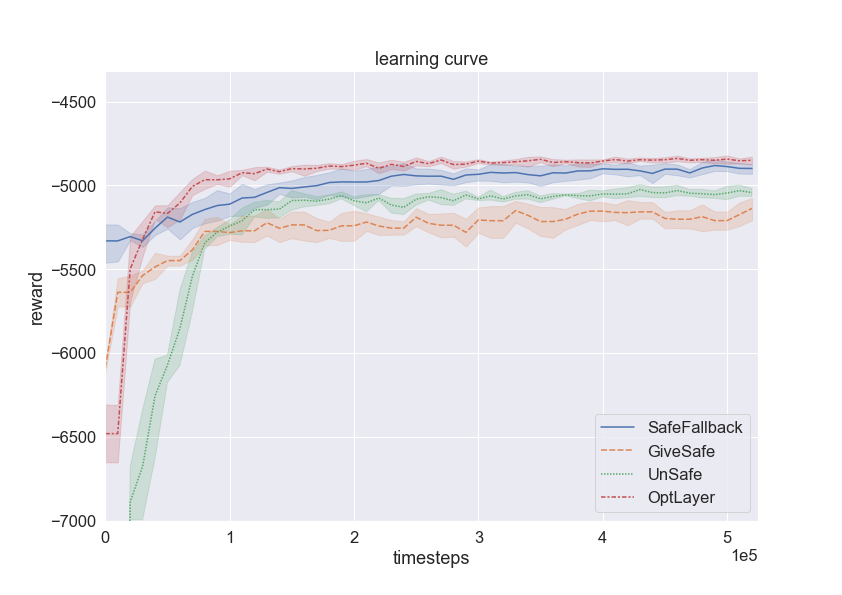}
    \caption{5-run average learning curves with a training budget of 15-years worth of time steps per run (i.e. 525.150 time steps per run)}
    \label{fig: learning curve}
\end{figure}

\begin{figure}[H]
    \centering
    \includegraphics[width=0.90\textwidth, height=0.4\textheight]{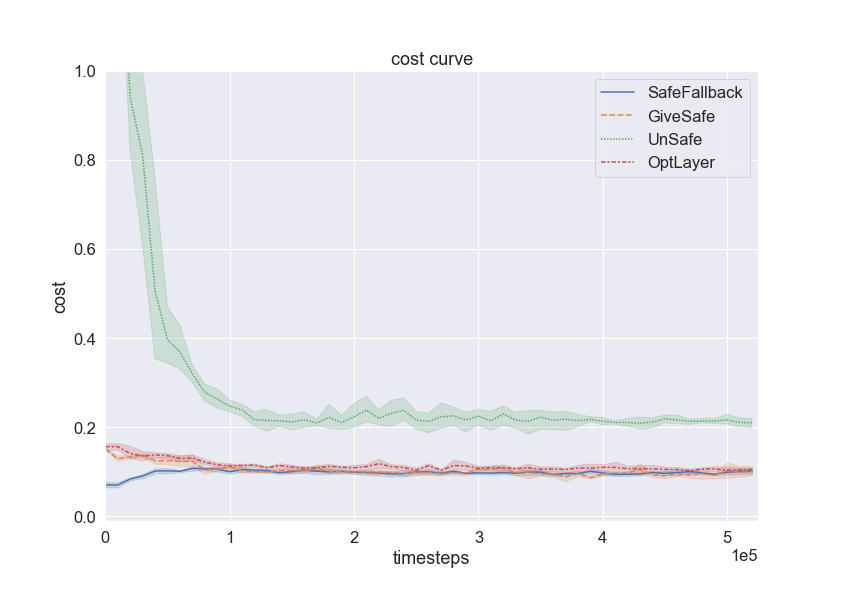}
    \caption{5-run average cost curve (i.e. constraint tolerance) with a training budget of 15-years worth of time steps per run (i.e. 525.150 time steps per run)}
    \label{fig: cost curve}
\end{figure}

\par The learning curves of the TD3 agents (using \autoref{algo: SafeFallback}, \autoref{algo: GiveSafe} and OptLayer, as well as without any safety layer - indicated as UnSafe) are presented in \autoref{fig: learning curve}, where the initial (at time step 0) and final (at time step 525.150) results are the figures reported in \autoref{tab: sim results}. We observe a steep initial learning rate, low variance, and a stable (slightly increasing) performance with increasing number of interactions with its environment. We also observe that \autoref{algo: SafeFallback} (SafeFallback), \autoref{algo: GiveSafe} (GiveSafe) and OptLayer have a significantly higher initial performance (before any training has occurred, i.e. at time step 0) compared to its \textit{vanilla} unsafe TD3 counterpart. This, again, due to the \textit{a priori} expert knowledge in the form of a known safe fallback policy and in the form of safety constraint equations. The unsafe RL agent only reaches the \textit{initial} performance (-6.481) of OptLayer after \(\sim\) 35.000 time steps (1 year) and of \autoref{algo: GiveSafe} (-6.089) after \(\sim\) 50.000 time steps (1 year and 5 months) and of \autoref{algo: SafeFallback} (-5.331) after \(\sim\) 85.000 time steps (2 years and 5 months). Furthermore, the initial performance of Optlayer is lower then \autoref{algo: SafeFallback} and \autoref{algo: GiveSafe}, surpassing \autoref{algo: GiveSafe} after \(\sim\) 18.575 time steps (6 months) and \autoref{algo: SafeFallback} after \(\sim\) 30.000 time steps (10 months) and that the performance gap remains significant with \autoref{algo: GiveSafe} (5,8 \%) while \autoref{algo: SafeFallback} reaches a similar performance (1,1\%).
\par The cost curves (constraint tolerance) of the TD3 agents are presented in \autoref{fig: cost curve}. We observe a steep initial decrease of the constraint tolerance of the \textit{vanilla} TD3 agent, yet never converging to the safety threshold of  15\% as defined by \autoref{equation3c} - while \autoref{algo: SafeFallback} and \autoref{algo: GiveSafe} never exceed this threshold (i.e. proving the hard-constraint satisfaction during training) and OptLayer slightly exceeds this threshold as noted here before. The constraint tolerance convergence of all safe methods is less steep and reaches a stable performance (\(\sim 10\%\)) after approximately 3 years (\(\sim 1e^5\) time steps).
\par However, the performance in terms of both the utility (objective value) and cost (constraint tolerance) of \autoref{algo: SafeFallback} (SafeFallback), \autoref{algo: GiveSafe} (GiveSafe) and OptLayer relies on an accurate formulation of the actual constraints, i.e. the accurate formulation of \autoref{equation3c} in this case study. As presented in \autoref{chapter4.3}, this is not always trivial - especially for the TESS and the HP. When we replace the pre-trained TESS and HP functions from the safety layer, by simpler (linear) analytical equations we observe a reduction in performance (i.e. the equations also used in OptLayer). For example, for \autoref{algo: SafeFallback} by its initial objective value of -5.331 to -5.431 and its initial constraint tolerance of 7,0\% to 7,6\%. This problem can be mitigated however, by artificially lowering \(Q_{tol}\) from \autoref{equation3c}.
\par Nevertheless, training a RL agent on a real safety-critical environment would only be possible with a sufficiently accurate safety layer (i.e. constraints) and using adequate \textit{a priori} unknown hyper-parameters. Therefore, we propose the following directions for future work:

\begin{itemize}
    \item Providing chance-constraint satisfaction guarantees for when no constraint functions are available \textit{a priori} (but only limited, known to be safe, operational data) and to safely improve the \textit{a prior} constraint functions as more data becomes available. Exploratory and exploitative steps will then never leave the safe region with high probability, by updating a statistical model (e.g. a Gaussian Process model).
    \item Further reducing the training budget (i.e. improving sample efficiency, e.g. by combining SafeFallback with OptLayer) and rolling out a fixed sequence of (robust\footnotemark) control actions (e.g. using model-based RL agents) for day-ahead market \textit{planning}.
    \item Robustness of the RL-based energy management systems under faulty and noisy measurements / observations and utilizing \textit{online} hyper-parameter optimization methods (i.e. that the hyper-parameters are tuned during online training).
\end{itemize}
\footnotetext{The transition probability matrix can then also be used to generate a \textit{robust} planning rather than a pure \textit{most-likelihood} planning} 

\section{Conclusion} \label{chapter6}
\par This paper presented two novel model-free safe RL methods, where the safety constraint formulation is decoupled from the RL (MDP) formulation. These provide hard-constraint, rather than soft- and chance-constraint, satisfaction guarantees both during training a (near) optimal policy (which involves exploratory and exploitative, i.e. greedy, steps) as well as during deployment of any policy (e.g. random agents or offline trained RL agents). These methods are demonstrated in a multi-energy management systems context, where detailed simulation results are provided. 
\par Both of the proposed methods are viable safety constraint handling techniques applicable beyond state-of-the-art RL, as demonstrated by random agents while still providing strict safety guarantees. Preferably, however, \autoref{algo: SafeFallback} (SafeFallback) is used as it showed good performance, does not requires to solve a mathematical program (e.g. a mixed-integer quadratic program in the case of OptLayer), and as the availability of a simple safe fallback policy is common or relatively easily constructible (i.e. in the form of a simple rule-based policy, e.g. a priority-based control strategy).

\section{Acknowledgement}
This work has been supported in part by ABB n.v. and Flemish Agency for Innovation and Entrepreneurship (VLAIO) grant HBC.2019.2613.

\section*{CRediT authorship contribution statement}
\par \textbf{Glenn Ceusters}: Conceptualization, Methodology, Software, Validation, Formal analysis, Resources, Data curation, Writing - original draft, Visualization, Funding acquisition; \textbf{Luis Ramirez Camargo}: Conceptualization, Writing - review and editing, Supervision; \textbf{Rüdiger Franke}: Supervision;  \textbf{Ann Nowé}: Writing - review and editing, Supervision; \textbf{Maarten Messagie}: Supervision. 

\appendix
\section{Simulations visualization}
\label{Appendix A}
\par
\begin{figure}[H]
    \centering
    \includegraphics[width=\textwidth,height=\textheight,keepaspectratio]{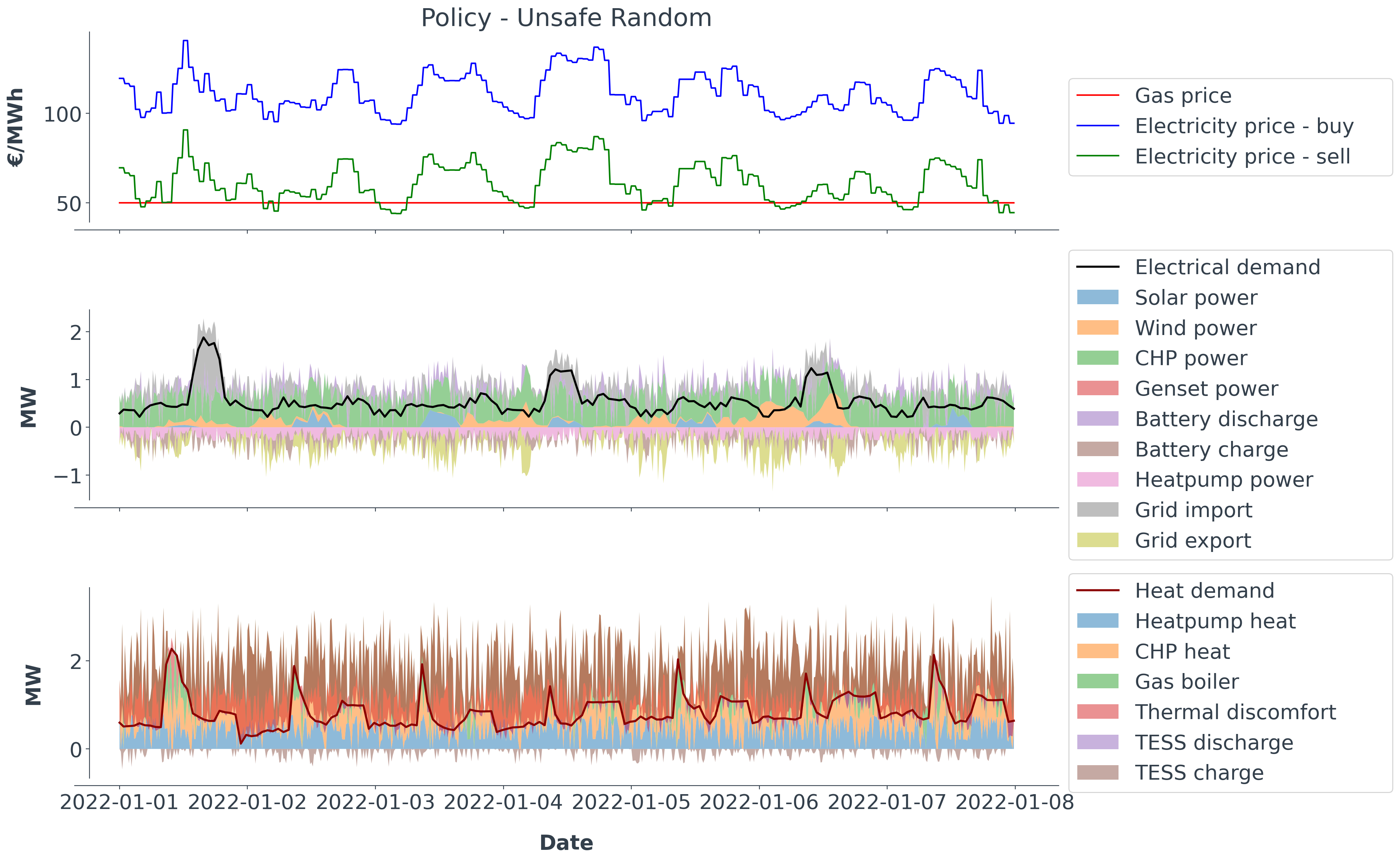}
    \caption{Policy visualization: unsafe random (or TD3 before training)}
    \label{fig: policy unsafe random}
\end{figure}

\begin{figure}[H]
    \centering
    \includegraphics[width=\textwidth,height=\textheight,keepaspectratio]{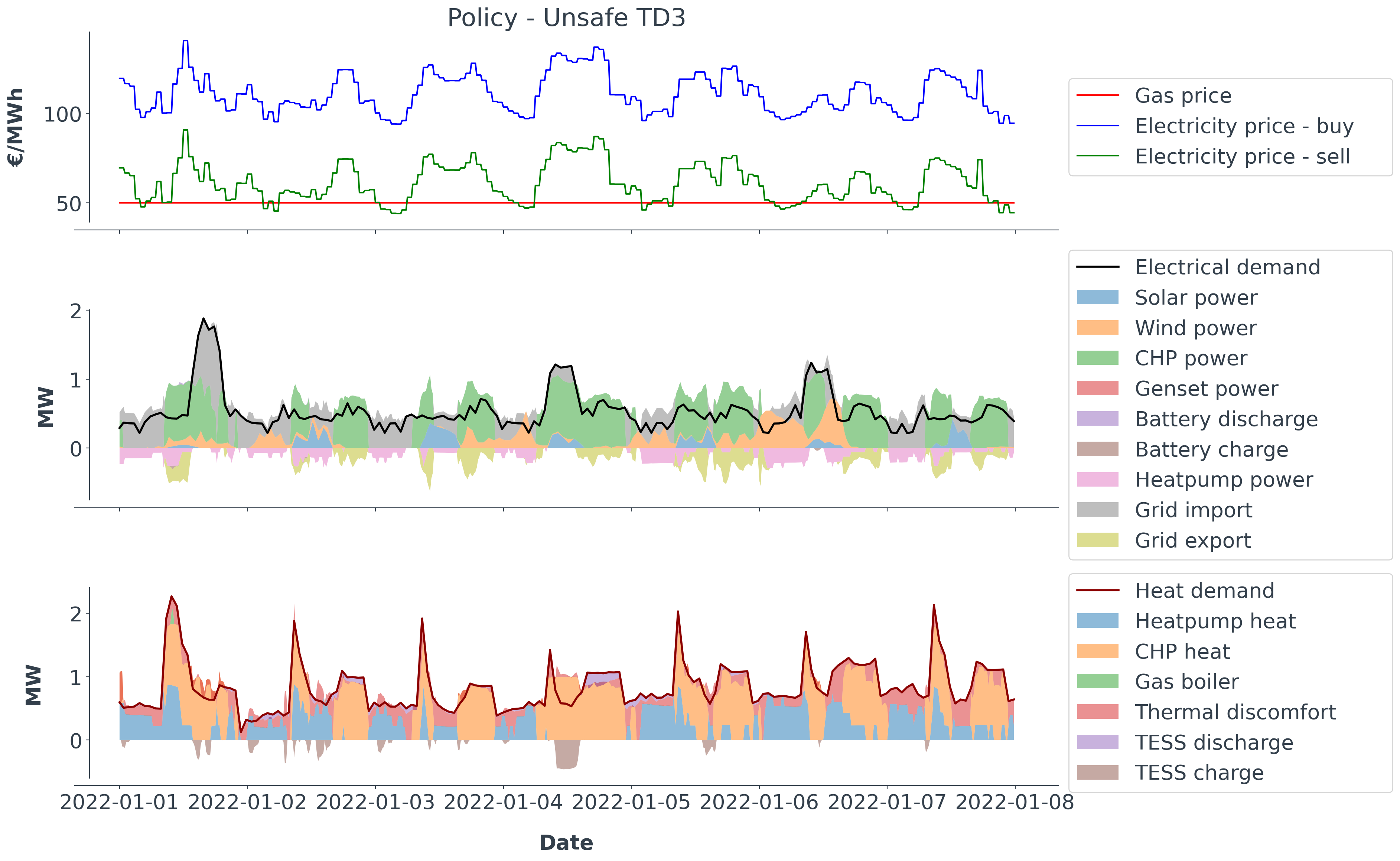}
    \caption{Policy visualization: unsafe TD3 (after unsafe training)}
    \label{fig: policy unsafe TD3}
\end{figure}

 In this section, we show a time series visualisation sample (a week) of the found control policies. The first observation that can be made (in \autoref{fig: policy unsafe random}) is the violently unsafe behavior (146,0\% of constraint tolerance, \autoref{tab: sim results}) of the TD3 agent before training, which at this stage acts as an unconstrained random agent. Specifically at this stage, a large thermal overproduction is the cause of the thermal discomfort and thus the constraint violation (as the total thermal installed capacity, given in \autoref{tab: mes dimensions}, is significantly higher then the thermal demand, e.g. due to the back-up boiler capacity - and given the random behaviour before training, the sum of the total thermal output is expected to be significantly high). The overproduction is avoided after training the TD3 agent (\autoref{fig: policy unsafe TD3}). The policy itself has a high utility, yet now a significant thermal underproduction is observed (21,0\% of constraint tolerance, \autoref{tab: sim results}). In practice, the natural gas boiler could be forced on to satisfy the thermal underproduction (yet this by itself would be an \textit{a prior} "fallback" policy).

\par
\begin{figure}[H]
    \centering
    \includegraphics[width=\textwidth,height=\textheight,keepaspectratio]{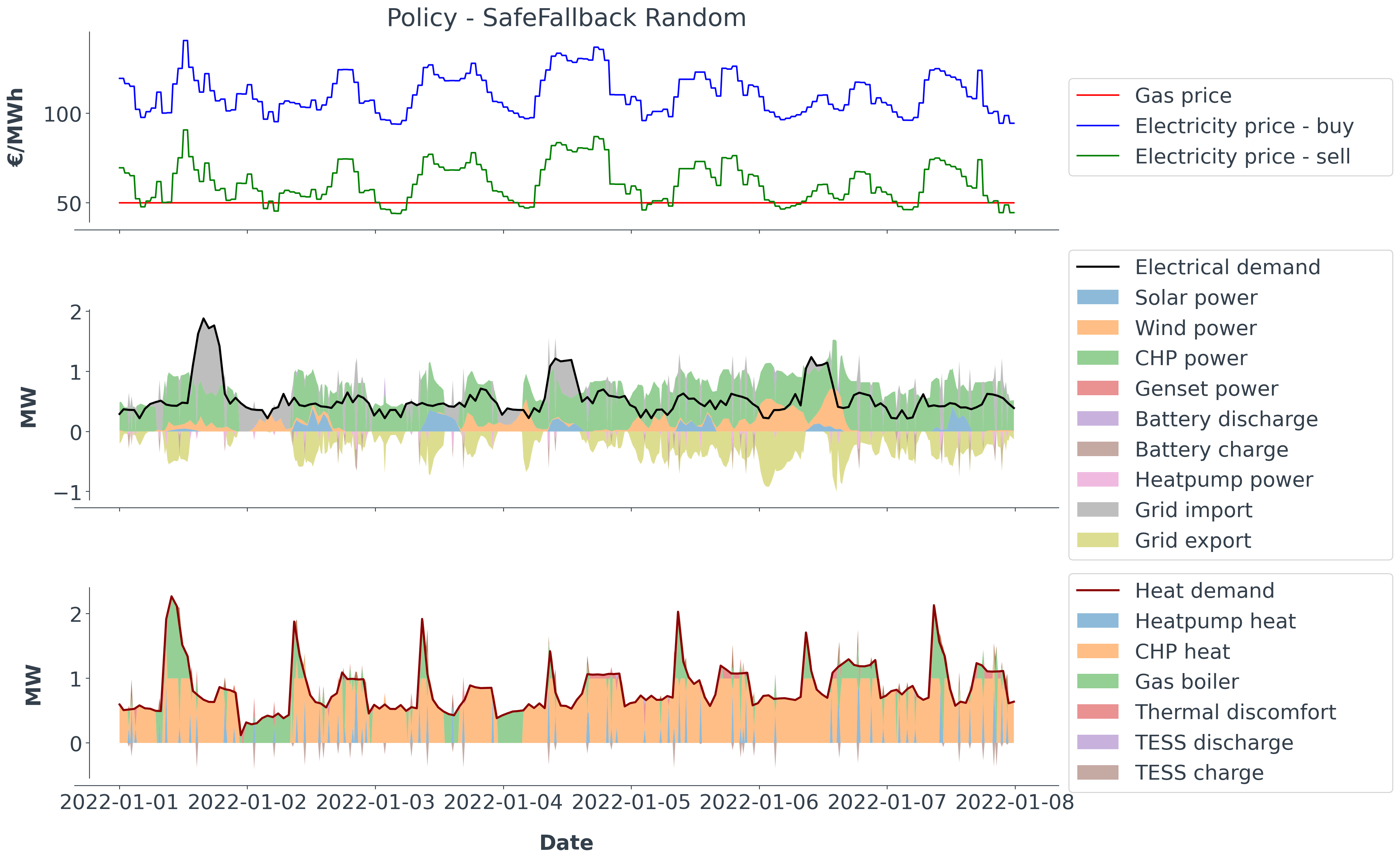}
    \caption{Policy visualization: SafeFallback random (or TD3 before training)}
    \label{fig: policy safefallback random}
\end{figure}

\begin{figure}[H]
    \centering
    \includegraphics[width=\textwidth,height=\textheight,keepaspectratio]{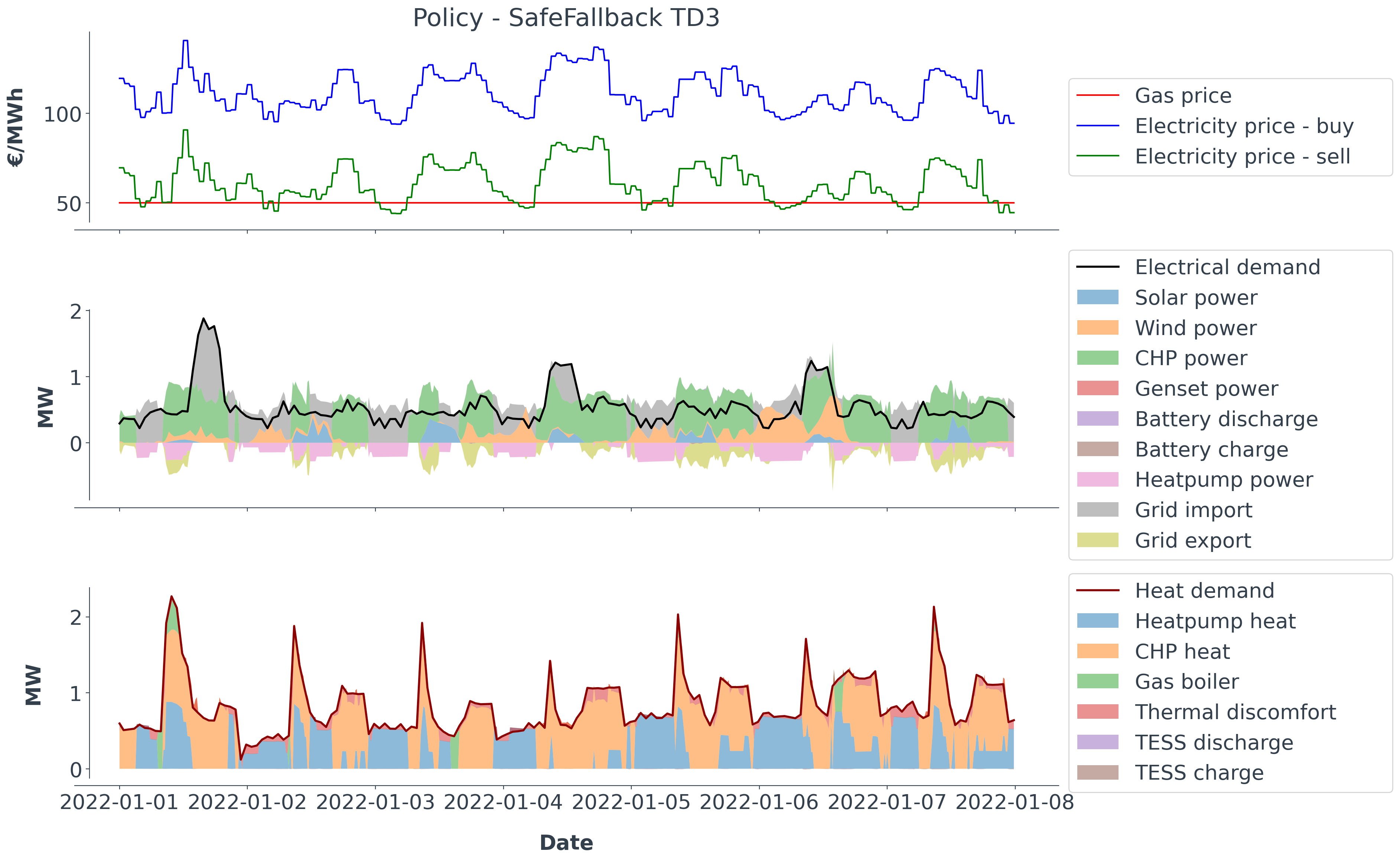}
    \caption{Policy visualization: SafeFallback TD3 (after safe training)}
    \label{fig: policy safefallback TD3}
\end{figure}

 When analysing the SafeFallback (\autoref{algo: SafeFallback}) policies, and comparing them against the \textit{vanilla} unsafe TD3 policies, we observe safe behavior. Before training, the constraint check mostly fails - using the safe fallback policy. Initially (\autoref{fig: policy safefallback random}), when the constraint check passes, safe random actions are observed (e.g. thermal "overproduction" is properly stored in the TESS). After \textit{safely} training the TD3 agent, a policy with a high utility and a low constraint tolerance is observed (\autoref{fig: policy safefallback TD3}). Thermal underproduction is still present, yet within the set bound \(Q_{tol}\) from \autoref{equation3c}.    

\par
\begin{figure}
    \centering
    \includegraphics[width=\textwidth,height=\textheight,keepaspectratio]{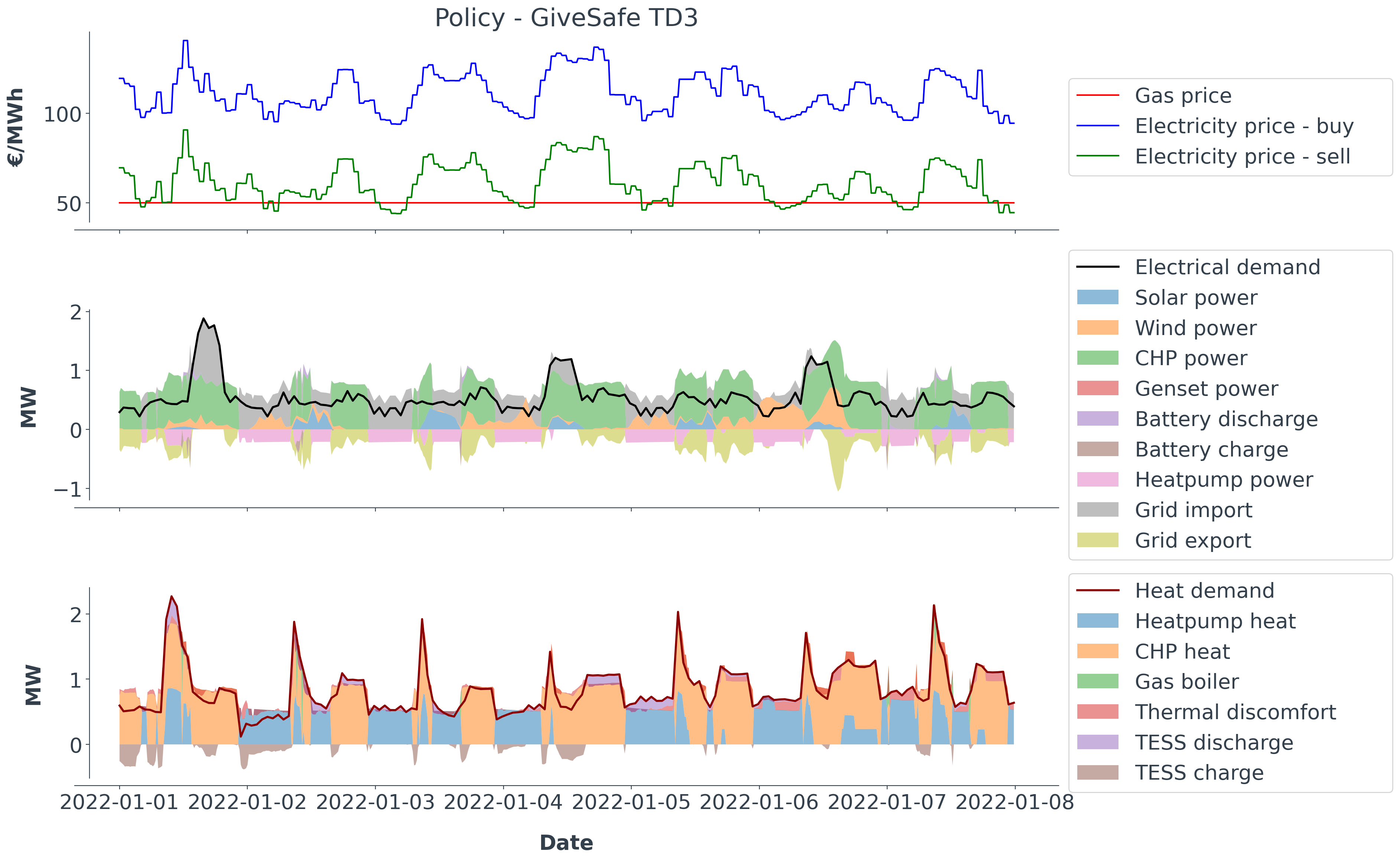}
    \caption{Policy visualization: GiveSafe TD3 (after safe training)}
    \label{fig: policy givesafe TD3}
\end{figure}

 When analysing the GiveSafe (\autoref{algo: GiveSafe}) policies, and comparing them against the \textit{vanilla} unsafe TD3 policies, we again observe safe behavior. Before training, \textit{all} actions are random but safe (e.g. thermal demand is matched by the thermal production or any thermal "overproduction" is stored in the TESS) - resulting in both a lower initial utility and higher constraint tolerance as \autoref{fig: policy safefallback random}. After \textit{safely} training the TD3 agent, a policy with a high utility and a low constraint tolerance is observed (\autoref{fig: policy givesafe TD3}) - yet with a lower utility as \autoref{fig: policy safefallback TD3}. Thermal underproduction is again still present, yet within the set bound \(Q_{tol}\) from \autoref{equation3c}.

\begin{figure}[H]
    \centering
    \includegraphics[width=\textwidth,height=\textheight,keepaspectratio]{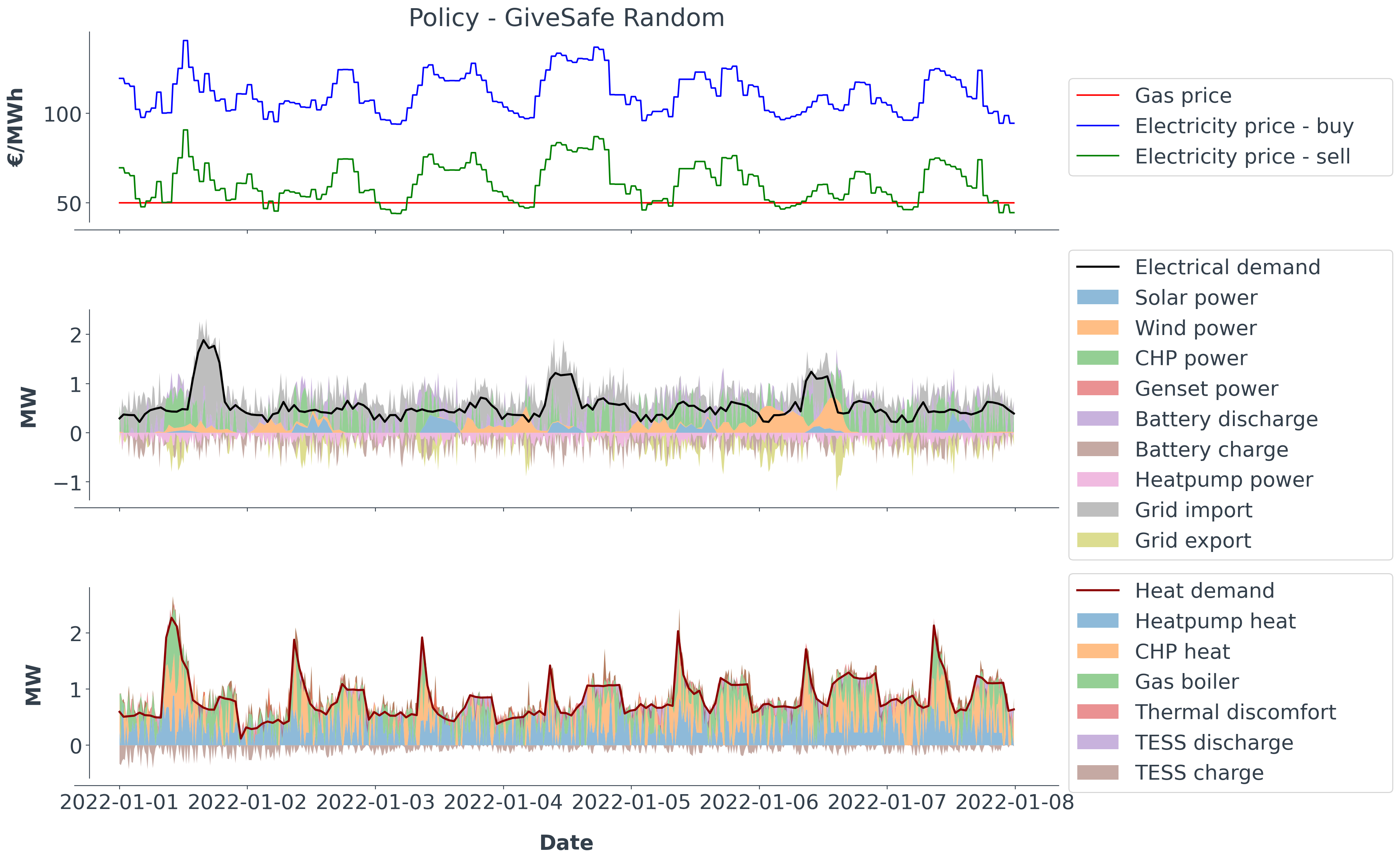}
    \caption{Policy visualization: GiveSafe random (or TD3 before training)}
    \label{fig: policy givesafe random}
\end{figure}

\begin{figure}[H]
    \centering
    \includegraphics[width=\textwidth,height=\textheight,keepaspectratio]{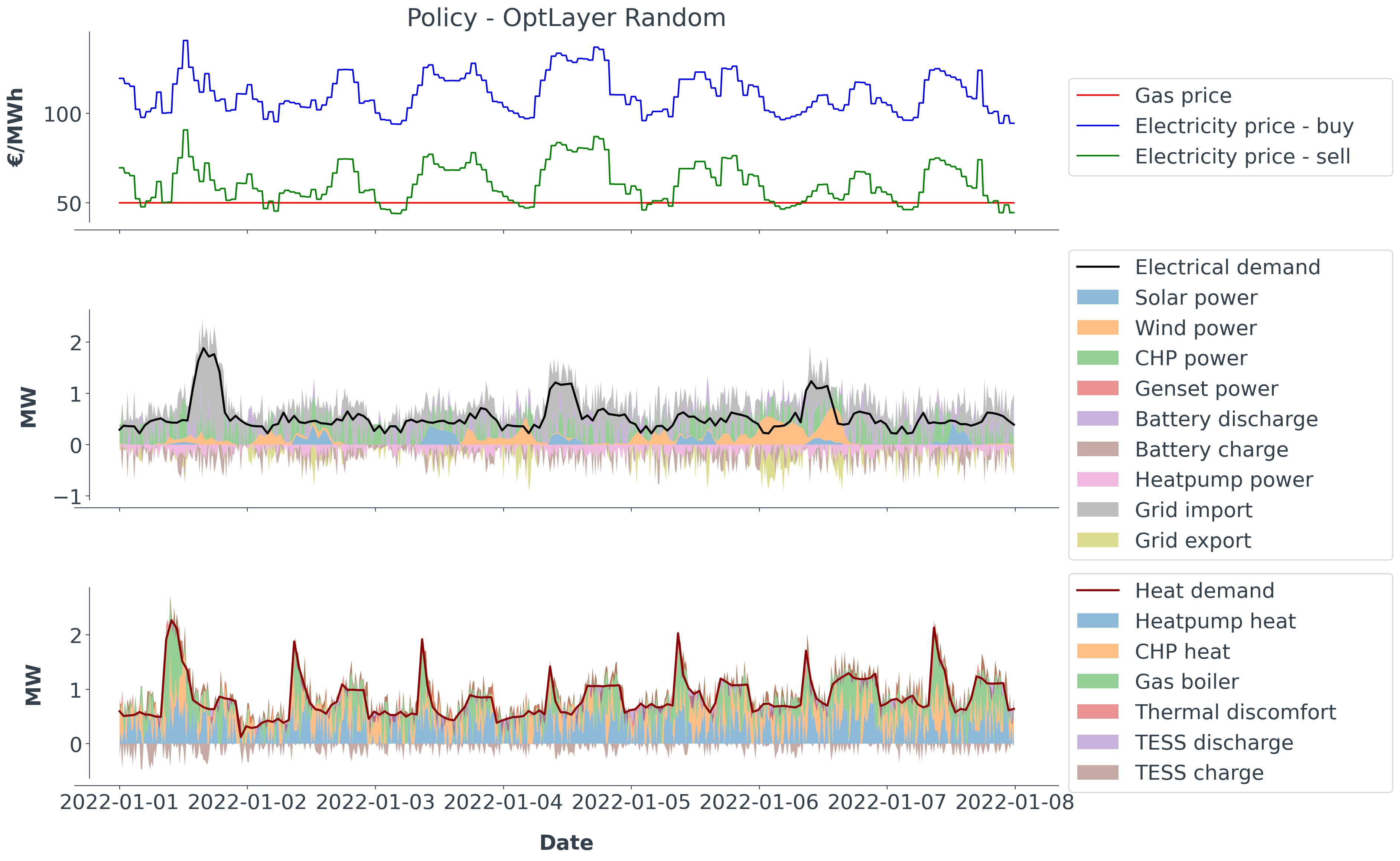}
    \caption{Policy visualization: OptSafe random (or TD3 before training)}
    \label{fig: policy optsafe random}
\end{figure}

\begin{figure}
    \centering
    \includegraphics[width=\textwidth,height=\textheight,keepaspectratio]{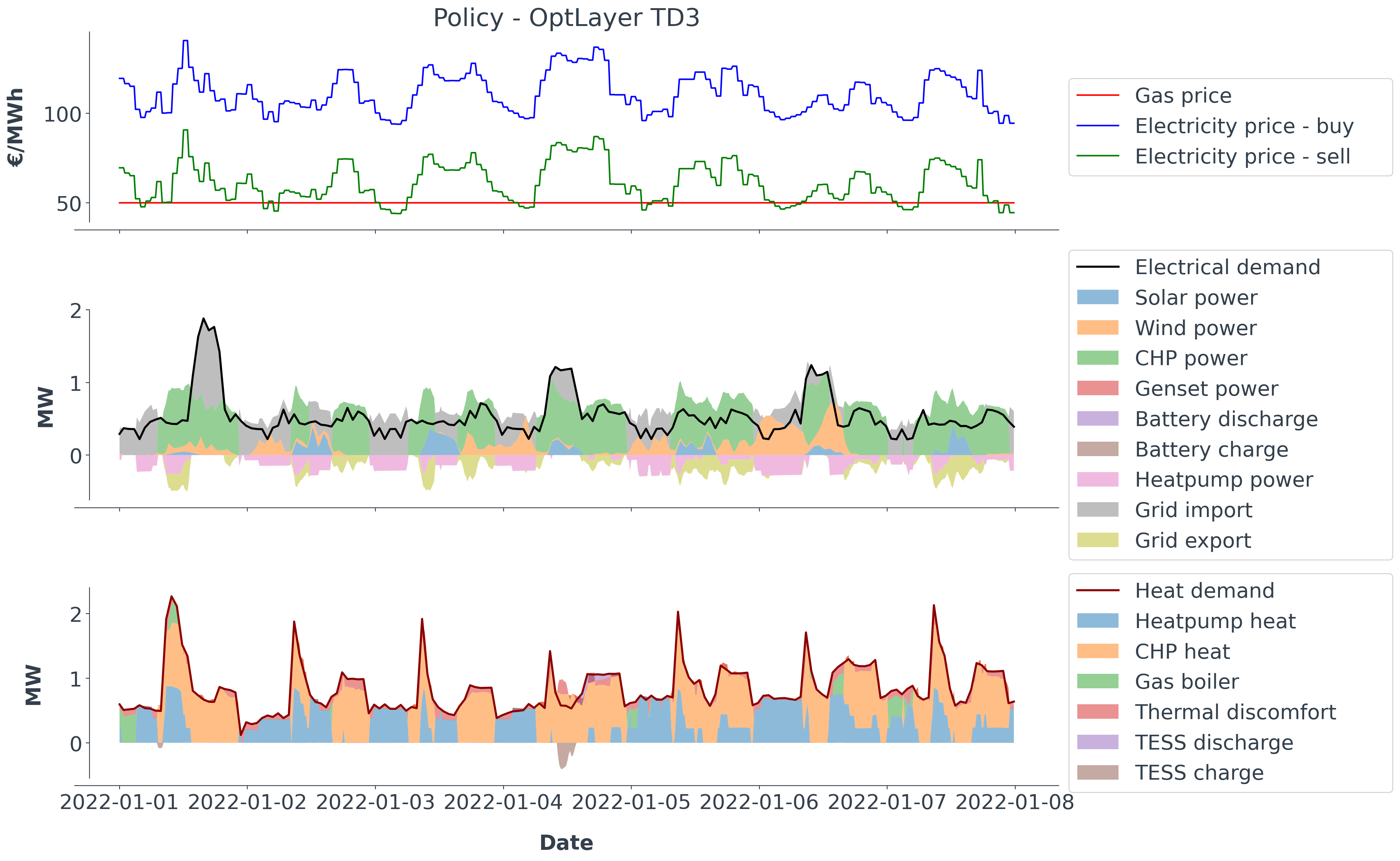}
    \caption{Policy visualization: OptLayer TD3 (after safe training)}
    \label{fig: policy optlayer TD3}
\end{figure}

Finally, when analysing the OptLayer policies, we again observe safe behavior (even though the maximum tolerance of 15\% is slightly violated, i.e. 15,6\% as discussed before). Before training (\autoref{fig: policy optsafe random}), all the actions proposed by the TD3 agents are random and are therefore corrected towards the closed feasible actions (see \cite{Pham2018OptLayerWorld} for the details of this algorithm). Even though this resembles the policy from \autoref{algo: GiveSafe}, these actions are then no longer completely random and almost always result in a distribution among actions (every continuous action all have some part in the feasible solution) and this resulting in a worse initial utility. After \textit{safely} training the TD3 agent (\autoref{fig: policy optlayer TD3}), a policy with a high utility and a low constraint tolerance is observed, yet again with some thermal underproduction (within the inequality bounds) as expected due to the conflicting objectives (the first term in \autoref{equation4g} minimizes the energy costs and therefor the production). 

\section{Pseudo-code of TD3}
\label{Appendix B}
\begin{algorithm}[H]
\DontPrintSemicolon
\SetAlgoLined
 \nl Input: initial policy parameters \( \theta \), Q-function parameters \( \phi_1 \), \( \phi_2 \), \\ 
 empty replay buffer \(\mathcal{D}\) \;
 \nl Set target parameters equal to main parameters \( \theta_{targ} \leftarrow \theta\), \( \theta_{targ,1} \leftarrow \theta_1\), \( \theta_{targ,2} \leftarrow \theta_2\) \;
 \nl \textbf{repeat} \;
 \nl \hspace{0.2cm} Observe state \(s\) and select action \(a = \text{clip}(\mu_{\theta}(s) + \epsilon, a_{Low}, a_{High}) \), where \( \epsilon \sim \mathcal{N} \) \;
 \nl \hspace{0.2cm} Execute \( a \) in the environment \;
 \nl \hspace{0.2cm} Observe next state \( s' \), reward \( r \) and done signal \( d \) to indicate whether \( s' \) is terminal \;
 \nl \hspace{0.2cm} Store \( (s,a,r,s',d) \) in replay buffer \( \mathcal{D} \) \;
 \nl \hspace{0.2cm} If \( s' \) is terminal, reset environment state \;
 \nl \hspace{0.2cm} \textbf{if} it's time to update \textbf{then} \;
 \nl \hspace{0.4cm} \textbf{for} \( j \) in range(however many updates) \textbf{do} \;
 \nl \hspace{0.6cm} Randomly sample a batch of transitions, \( B = \{ (s,a,r,s',d) \} \) from \( \mathcal{D} \) \;
 \nl \hspace{0.6cm} Compute target actions
 \begin{align*}
     a'(s') = \text{clip}\big( \mu_{\theta_{targ}}(s') + \text{clip}(\epsilon, -c, c), a_{Low}, a_{High} \big),& & \epsilon \sim \mathcal{N}(0,\sigma)
 \end{align*} \vspace{-0.5cm}\;
 \nl \hspace{0.6cm} Compute targets
 \begin{equation*}
     y(r,s',d) = r + \gamma(1 - d) \min_{i=1,2} Q_{\phi_{targ,i}}(s',a'(s'))
 \end{equation*} \vspace{-0.5cm}\;
 \nl \hspace{0.6cm} Update Q-function by one step of gradient descent using
 \begin{align*}
     \nabla_{\phi_{i}} \frac{1}{|B|} \sum_{(s,a,r,s',d)\in B}(Q_{\phi_{i}}(s,a) - y(r,s',d))^2& & \text{for} \ \ i=1,2
 \end{align*} \vspace{-0.5cm}\;
 \nl \hspace{0.6cm} \textbf{if} \( \ j \ \) mod \(\ \texttt{policy\_delay} = 0\) \textbf{then} \;
 \nl \hspace{0.8cm} Update policy by one step of gradient ascent using
 \begin{equation*}
     \nabla_\theta \frac{1}{|B|} \sum_{s\in B} Q_{\phi_{1}}(s,\mu_\theta(s))
 \end{equation*}  \vspace{-0.5cm}\;
 \nl \hspace{0.8cm} Update target networks with
 \begin{align*}
     \phi_{targ,i} \leftarrow \rho \phi_{targ,i} + (1 - \rho) \phi_i& & \text{for} \ \ i=1,2 \\
     \theta_{targ} \leftarrow \rho \theta_{targ} + (1 - \rho) \theta
 \end{align*} \vspace{-0.5cm} \;
 \nl \hspace{0.6cm} \textbf{end if} \;
 \nl \hspace{0.4cm} \textbf{end for} \;
 \nl \hspace{0.2cm} \textbf{end if} \;
 \nl \textbf{until} convergence
 \caption{Twin Delayed DDPG (TD3) \cite{OpenAI2020TwinDocumentation}}\label{algo: td3}
\end{algorithm}

\section{Run-time statistics}
\label{Appendix C}

\par The simulations are conducted on a local machine with a Intel® Core™ i5-8365U CPU @1.6GHz, 16 GB of Ram and an SSD. Over a yearly simulation, the following run-time statistics per simulated time-step (with a control horizon of 15 min) are observed.

\begin{table}[!h]
    \centering
    \begin{tabular}{l|c|c|c|c|c}
    \rowcolor[HTML]{efefef} 
    \textbf{Optimal controller} & \textbf{min} & \textbf{mean} & \textbf{std} & \textbf{max} & \textbf{total}\\
    \hline
    Unsafe TD3 & 0,027 s & 0,041 s & 0,006 s & 0,100 s & 1.424 s \\
    Unsafe Random & 0,027 s & 0,040 s & 0,011 s & 0,120 s & 1.394 s \\
    OptLayer TD3 & 0,044 s & 0,069 s & 0,031 s & 0,398 s & 2.418 s \\
    OptLayer Random & 0,041 s & 0,231 s & 0,084 s & 6,395 s & 8.091 s \\
    SafeFallback TD3 & 0,042 s & 0,058 s & 0,008 s & 0,250 s & 2.047 s \\ 
    SafeFallback Random & 0,037 s & 0,044 s & 0,005 s & 0,142 s & 1.545 s \\
    GiveSafe TD3 & 0,041 s & 0,060 s & 0,040 s & 2,661 s & 2.090 s \\
    GiveSafe Random & 0,041 s & 2,272 s & 3,760 s & 52,390 s & 79.615 s
    \end{tabular}
    \caption{Run-time statistics}
    \label{tab: run-time}
\end{table}

\par The maximum run-time per time-step never exceeds the control horizon of 15 minutes, as this otherwise would be considered impractical with the given hardware. We observe that the unsafe agents have the fastest run-time, as they don't have the constraint check to compute. Yet, we have argued that using unsafe agents is not realistic in safety-critical environments and are given for completeness only. Furthermore we observe that the \hyperref[algo: SafeFallback]{SafeFallback} method itself (demonstrated by using random agents) is significantly faster then the \hyperref[algo: GiveSafe]{GiveSafe} method, as the GiveSafe method can require multiple additional "offline" time-steps for every "online" (i.e. real) time-step, and is significantly faster than OptLayer, as this involves solving a mixed-integer quadratic problem (MIQP) for every time step an infeasible action is selected by the TD3 agent. After the TD3 agents are trained though, the run-time is approximately the same - as the amount of unsafe actions proposed by the TD3 agent (and thus the need for additional "offline" training steps or MIQP solving) is greatly reduced.
\par Notice that this are the run-time statistics \textit{after} training (i.e., pure policy execution, in the case of the TD3 agents). Including the online training run-time statistics (i.e. fitting the function approximation algorithm - which is a multi-layer perceptron in our case), the mean run-time would result in 0,058 s for the unsafe TD3 agent, 0,122 s for the OptLayer TD3 agent, 0,076 s for the SafeFallback TD3 agent and 2,229 s for the GiveSafe TD3 agent. These online training run-time statistics are still magnitudes faster then the control horizon of 15 minutes.

\reftitle{References}

\externalbibliography{yes}
\bibliography{references}

\begin{thebibliography}{-------}
\providecommand{\natexlab}[1]{#1}

\bibitem[Fabrizio \em{et~al.}(2009)Fabrizio, Filippi, and
  Virgone]{Fabrizio2009Trade-offSystems}
Fabrizio, E.; Filippi, M.; Virgone, J.
\newblock {Trade-off between environmental and economic objectives in the
  optimization of multi-energy systems}.
\newblock {\em Building Simulation 2009 2:1} {\bf 2009}, {\em 2},~29--40.
\newblock
  doi:{\changeurlcolor{black}\href{https://doi.org/10.1007/S12273-009-9202-4}{\detokenize{10.1007/S12273-009-9202-4}}}.

\bibitem[Engell(2007)]{Engell2007FeedbackOperation}
Engell, S.
\newblock {Feedback control for optimal process operation}.
\newblock {\em Journal of Process Control} {\bf 2007}, {\em 17},~203--219.
\newblock
  doi:{\changeurlcolor{black}\href{https://doi.org/10.1016/J.JPROCONT.2006.10.011}{\detokenize{10.1016/J.JPROCONT.2006.10.011}}}.

\bibitem[Ceusters \em{et~al.}(2021)Ceusters, Rodr{\'{i}}guez, Garc{\'{i}}a,
  Franke, Deconinck, Helsen, Now{\'{e}}, Messagie, and
  Camargo]{Ceusters2021Model-predictiveStudies}
Ceusters, G.; Rodr{\'{i}}guez, R.C.; Garc{\'{i}}a, A.B.; Franke, R.; Deconinck,
  G.; Helsen, L.; Now{\'{e}}, A.; Messagie, M.; Camargo, L.R.
\newblock {Model-predictive control and reinforcement learning in multi-energy
  system case studies}.
\newblock {\em Applied Energy} {\bf 2021}, {\em 303},~117634.
\newblock
  doi:{\changeurlcolor{black}\href{https://doi.org/10.1016/j.apenergy.2021.117634}{\detokenize{10.1016/j.apenergy.2021.117634}}}.

\bibitem[Cao \em{et~al.}(2020)Cao, Hu, Zhao, Zhang, Zhang, Liu, Chen, and
  Blaabjerg]{Cao2020ReinforcementReview}
Cao, D.; Hu, W.; Zhao, J.; Zhang, G.; Zhang, B.; Liu, Z.; Chen, Z.; Blaabjerg,
  F.
\newblock {Reinforcement Learning and Its Applications in Modern Power and
  Energy Systems: A Review}.
\newblock {\em Journal of Modern Power Systems and Clean Energy} {\bf 2020},
  {\em 8},~1029--1042.
\newblock
  doi:{\changeurlcolor{black}\href{https://doi.org/10.35833/MPCE.2020.000552}{\detokenize{10.35833/MPCE.2020.000552}}}.

\bibitem[Yang \em{et~al.}(2020)Yang, Zhao, Li, and
  Zomaya]{Yang2020ReinforcementSurvey}
Yang, T.; Zhao, L.; Li, W.; Zomaya, A.Y.
\newblock {Reinforcement learning in sustainable energy and electric systems: a
  survey}.
\newblock {\em Annual Reviews in Control} {\bf 2020}, {\em 49},~145--163.
\newblock
  doi:{\changeurlcolor{black}\href{https://doi.org/10.1016/J.ARCONTROL.2020.03.001}{\detokenize{10.1016/J.ARCONTROL.2020.03.001}}}.

\bibitem[Perera and Kamalaruban(2021)]{Perera2021ApplicationsSystems}
Perera, A.T.; Kamalaruban, P.
\newblock {Applications of reinforcement learning in energy systems}.
\newblock {\em Renewable and Sustainable Energy Reviews} {\bf 2021}, {\em
  137},~110618.
\newblock
  doi:{\changeurlcolor{black}\href{https://doi.org/10.1016/J.RSER.2020.110618}{\detokenize{10.1016/J.RSER.2020.110618}}}.

\bibitem[Rayati \em{et~al.}(2015)Rayati, Sheikhi, and
  Ranjbar]{Rayati2015ApplyingGrid}
Rayati, M.; Sheikhi, A.; Ranjbar, A.M.
\newblock {Applying reinforcement learning method to optimize an Energy Hub
  operation in the smart grid}.
\newblock {\em 2015 IEEE Power and Energy Society Innovative Smart Grid
  Technologies Conference, ISGT 2015} {\bf 2015}.
\newblock
  doi:{\changeurlcolor{black}\href{https://doi.org/10.1109/ISGT.2015.7131906}{\detokenize{10.1109/ISGT.2015.7131906}}}.

\bibitem[Watkins(1989)]{watkins1989learning}
Watkins, C.J.C.H.
\newblock {Learning from delayed rewards}.
\newblock PhD thesis, King's College, Cambridge United Kingdom,  1989.

\bibitem[Sheikhi \em{et~al.}(2016)Sheikhi, Rayati, and
  Ranjbar]{Sheikhi2016DemandSystems}
Sheikhi, A.; Rayati, M.; Ranjbar, A.M.
\newblock {Demand side management for a residential customer in multi-energy
  systems}.
\newblock {\em Sustainable Cities and Society} {\bf 2016}, {\em 22},~63--77.
\newblock
  doi:{\changeurlcolor{black}\href{https://doi.org/10.1016/j.scs.2016.01.010}{\detokenize{10.1016/j.scs.2016.01.010}}}.

\bibitem[Mbuwir \em{et~al.}(2018)Mbuwir, Kaffash, and
  Deconinck]{Mbuwir2018BatteryLearning}
Mbuwir, B.V.; Kaffash, M.; Deconinck, G.
\newblock {Battery Scheduling in a Residential Multi-Carrier Energy System
  Using Reinforcement Learning}.
\newblock  2018 IEEE International Conference on Communications, Control, and
  Computing Technologies for Smart Grids, SmartGridComm 2018. Institute of
  Electrical and Electronics Engineers Inc.,  2018.
\newblock
  doi:{\changeurlcolor{black}\href{https://doi.org/10.1109/SmartGridComm.2018.8587412}{\detokenize{10.1109/SmartGridComm.2018.8587412}}}.

\bibitem[Wang \em{et~al.}(2019)Wang, Chen, Wu, DIng, Lou, and
  Liu]{Wang2019Bi-levelSystem}
Wang, X.; Chen, H.; Wu, J.; DIng, Y.; Lou, Q.; Liu, S.
\newblock {Bi-level Multi-agents Interactive Decision-making Model in Regional
  Integrated Energy System}.
\newblock  2019 3rd IEEE Conference on Energy Internet and Energy System
  Integration: Ubiquitous Energy Network Connecting Everything, EI2 2019.
  Institute of Electrical and Electronics Engineers Inc.,  2019, pp.
  2103--2108.
\newblock
  doi:{\changeurlcolor{black}\href{https://doi.org/10.1109/EI247390.2019.9061889}{\detokenize{10.1109/EI247390.2019.9061889}}}.

\bibitem[Ahrarinouri \em{et~al.}(2020)Ahrarinouri, Rastegar, and
  Seifi]{Ahrarinouri2020Multi-AgentBuildings}
Ahrarinouri, M.; Rastegar, M.; Seifi, A.R.
\newblock {Multi-Agent Reinforcement Learning for Energy Management in
  Residential Buildings}.
\newblock {\em IEEE Transactions on Industrial Informatics} {\bf 2020}, p.~1.
\newblock
  doi:{\changeurlcolor{black}\href{https://doi.org/10.1109/tii.2020.2977104}{\detokenize{10.1109/tii.2020.2977104}}}.

\bibitem[Ye \em{et~al.}(2020)Ye, Ye, Qiu, Wu, Strbac, and
  Ward]{Ye2020Model-FreeLearning}
Ye, Y.; Ye, Y.; Qiu, D.; Wu, X.; Strbac, G.; Ward, J.
\newblock {Model-Free Real-Time Autonomous Control for a Residential
  Multi-Energy System Using Deep Reinforcement Learning}.
\newblock {\em IEEE Transactions on Smart Grid} {\bf 2020}, {\em
  11},~3068--3082.
\newblock
  doi:{\changeurlcolor{black}\href{https://doi.org/10.1109/TSG.2020.2976771}{\detokenize{10.1109/TSG.2020.2976771}}}.

\bibitem[Lillicrap \em{et~al.}(2015)Lillicrap, Hunt, Pritzel, Heess, Erez,
  Tassa, Silver, and Wierstra]{Lillicrap2015ContinuousLearning}
Lillicrap, T.P.; Hunt, J.J.; Pritzel, A.; Heess, N.; Erez, T.; Tassa, Y.;
  Silver, D.; Wierstra, D.
\newblock {Continuous control with deep reinforcement learning}.
\newblock {\em 4th International Conference on Learning Representations, ICLR
  2016 - Conference Track Proceedings} {\bf 2015}.

\bibitem[Xu \em{et~al.}(2021)Xu, Han, Liu, Martinez-Garcia, and
  Wang]{Xu2021Multi-energyEvolution}
Xu, Z.; Han, G.; Liu, L.; Martinez-Garcia, M.; Wang, Z.
\newblock {Multi-energy scheduling of an industrial integrated energy system by
  reinforcement learning-based differential evolution}.
\newblock {\em IEEE Transactions on Green Communications and Networking} {\bf
  2021}, {\em 5},~1077--1090.
\newblock
  doi:{\changeurlcolor{black}\href{https://doi.org/10.1109/TGCN.2021.3061789}{\detokenize{10.1109/TGCN.2021.3061789}}}.

\bibitem[Zhu \em{et~al.}(2022)Zhu, Yang, Liu, Wang, Ma, and
  Guan]{Zhu2022EnergyPark}
Zhu, D.; Yang, B.; Liu, Y.; Wang, Z.; Ma, K.; Guan, X.
\newblock {Energy Management Based on Multi-Agent Deep Reinforcement Learning
  for A Multi-Energy Industrial Park}.
\newblock {\em Applied Energy} {\bf 2022}, {\em 311},~118636.
\newblock
  doi:{\changeurlcolor{black}\href{https://doi.org/10.1016/j.apenergy.2022.118636}{\detokenize{10.1016/j.apenergy.2022.118636}}}.

\bibitem[Pu \em{et~al.}(2021)Pu, Wang, Yang, Yao, and
  Li]{Pu2021DecomposedLearning}
Pu, Y.; Wang, S.; Yang, R.; Yao, X.; Li, B.
\newblock {Decomposed Soft Actor-Critic Method for Cooperative Multi-Agent
  Reinforcement Learning}.
\newblock {\em arXiv} {\bf 2021}.

\bibitem[Schulman \em{et~al.}(2017)Schulman, Wolski, Dhariwal, Radford, and
  Klimov]{Schulman2017ProximalAlgorithms}
Schulman, J.; Wolski, F.; Dhariwal, P.; Radford, A.; Klimov, O.
\newblock {Proximal Policy Optimization Algorithms}.
\newblock {\em arxiv} {\bf 2017}.

\bibitem[Fujimoto \em{et~al.}(2018)Fujimoto, Van~Hoof, and
  Meger]{Fujimoto2018AddressingMethods}
Fujimoto, S.; Van~Hoof, H.; Meger, D.
\newblock {Addressing Function Approximation Error in Actor-Critic Methods}.
\newblock {\em 35th International Conference on Machine Learning, ICML 2018}
  {\bf 2018}, {\em 4},~2587--2601.

\bibitem[Zhang \em{et~al.}(2021)Zhang, Hu, Cao, Li, Zhang, Chen, and
  Blaabjerg]{Zhang2021SoftEnergy}
Zhang, B.; Hu, W.; Cao, D.; Li, T.; Zhang, Z.; Chen, Z.; Blaabjerg, F.
\newblock {Soft actor-critic –based multi-objective optimized energy
  conversion and management strategy for integrated energy systems with
  renewable energy}.
\newblock {\em Energy Conversion and Management} {\bf 2021}, {\em 243},~114381.
\newblock
  doi:{\changeurlcolor{black}\href{https://doi.org/https://doi.org/10.1016/j.enconman.2021.114381}{\detokenize{https://doi.org/10.1016/j.enconman.2021.114381}}}.

\bibitem[Zhang \em{et~al.}(2019)Zhang, Hu, Cao, Huang, Chen, and
  Blaabjerg]{Zhang2019DeepEnergy}
Zhang, B.; Hu, W.; Cao, D.; Huang, Q.; Chen, Z.; Blaabjerg, F.
\newblock {Deep reinforcement learning–based approach for optimizing energy
  conversion in integrated electrical and heating system with renewable
  energy}.
\newblock {\em Energy Conversion and Management} {\bf 2019}, {\em 202},~112199.
\newblock
  doi:{\changeurlcolor{black}\href{https://doi.org/https://doi.org/10.1016/j.enconman.2019.112199}{\detokenize{https://doi.org/10.1016/j.enconman.2019.112199}}}.

\bibitem[Zhang \em{et~al.}(2020)Zhang, Hu, Li, Cao, Huang, Huang, Chen, and
  Blaabjerg]{Zhang2020DynamicApproach}
Zhang, B.; Hu, W.; Li, J.; Cao, D.; Huang, R.; Huang, Q.; Chen, Z.; Blaabjerg,
  F.
\newblock {Dynamic energy conversion and management strategy for an integrated
  electricity and natural gas system with renewable energy: Deep reinforcement
  learning approach}.
\newblock {\em Energy Conversion and Management} {\bf 2020}, {\em 220},~113063.
\newblock
  doi:{\changeurlcolor{black}\href{https://doi.org/https://doi.org/10.1016/j.enconman.2020.113063}{\detokenize{https://doi.org/10.1016/j.enconman.2020.113063}}}.

\bibitem[Zhang \em{et~al.}(2022)Zhang, Hu, Cao, Zhang, Huang, Chen, and
  Blaabjerg]{Zhang2022AFreshwater}
Zhang, G.; Hu, W.; Cao, D.; Zhang, Z.; Huang, Q.; Chen, Z.; Blaabjerg, F.
\newblock {A multi-agent deep reinforcement learning approach enabled
  distributed energy management schedule for the coordinate control of
  multi-energy hub with gas, electricity, and freshwater}.
\newblock {\em Energy Conversion and Management} {\bf 2022}, {\em 255},~115340.
\newblock
  doi:{\changeurlcolor{black}\href{https://doi.org/https://doi.org/10.1016/j.enconman.2022.115340}{\detokenize{https://doi.org/10.1016/j.enconman.2022.115340}}}.

\bibitem[Venayagamoorthy \em{et~al.}(2016)Venayagamoorthy, Sharma, Gautam, and
  Ahmadi]{Venayagamoorthy2016DynamicMicrogrid}
Venayagamoorthy, G.K.; Sharma, R.K.; Gautam, P.K.; Ahmadi, A.
\newblock {Dynamic Energy Management System for a Smart Microgrid}.
\newblock {\em IEEE Transactions on Neural Networks and Learning Systems} {\bf
  2016}, {\em 27},~1643--1656.
\newblock
  doi:{\changeurlcolor{black}\href{https://doi.org/10.1109/TNNLS.2016.2514358}{\detokenize{10.1109/TNNLS.2016.2514358}}}.

\bibitem[Zhang \em{et~al.}(2020)Zhang, Dehghanpour, Wang, and
  Huang]{Zhang2020AInformation}
Zhang, Q.; Dehghanpour, K.; Wang, Z.; Huang, Q.
\newblock {A Learning-Based Power Management Method for Networked Microgrids
  under Incomplete Information}.
\newblock {\em IEEE Transactions on Smart Grid} {\bf 2020}, {\em
  11},~1193--1204.
\newblock
  doi:{\changeurlcolor{black}\href{https://doi.org/10.1109/TSG.2019.2933502}{\detokenize{10.1109/TSG.2019.2933502}}}.

\bibitem[Zhao \em{et~al.}(2020)Zhao, Zhao, Qiu, Liang, and
  Dong]{Zhao2020CooperativeLearning}
Zhao, H.; Zhao, J.; Qiu, J.; Liang, G.; Dong, Z.Y.
\newblock {Cooperative Wind Farm Control with Deep Reinforcement Learning and
  Knowledge-Assisted Learning}.
\newblock {\em IEEE Transactions on Industrial Informatics} {\bf 2020}, {\em
  16},~6912--6921.
\newblock
  doi:{\changeurlcolor{black}\href{https://doi.org/10.1109/TII.2020.2974037}{\detokenize{10.1109/TII.2020.2974037}}}.

\bibitem[Park \em{et~al.}(2022)Park, Min, Ryu, and
  Choi]{Park2022DIP-QL:Systems}
Park, H.; Min, D.; Ryu, J.h.; Choi, D.G.
\newblock {DIP-QL: A Novel Reinforcement Learning Method for Constrained
  Industrial Systems}.
\newblock {\em IEEE Transactions on Industrial Informatics} {\bf 2022}.
\newblock
  doi:{\changeurlcolor{black}\href{https://doi.org/10.1109/TII.2022.3159570}{\detokenize{10.1109/TII.2022.3159570}}}.

\bibitem[Pham \em{et~al.}(2018)Pham, De~Magistris, and
  Tachibana]{Pham2018OptLayerWorld}
Pham, T.H.; De~Magistris, G.; Tachibana, R.
\newblock {OptLayer - Practical Constrained Optimization for Deep Reinforcement
  Learning in the Real World}.
\newblock {\em Proceedings - IEEE International Conference on Robotics and
  Automation} {\bf 2018}, pp. 6236--6243.
\newblock
  doi:{\changeurlcolor{black}\href{https://doi.org/10.1109/ICRA.2018.8460547}{\detokenize{10.1109/ICRA.2018.8460547}}}.

\bibitem[Garc{\'{i}}a and Fern{\'{a}}ndez(2015)]{Garcia2015ALearning}
Garc{\'{i}}a, J.; Fern{\'{a}}ndez, F.
\newblock {A comprehensive survey on safe reinforcement learning},  2015.

\bibitem[Dulac-Arnold \em{et~al.}(2021)Dulac-Arnold, Levine, Mankowitz, Li,
  Paduraru, Gowal, and Hester]{Dulac-Arnold2021ChallengesAnalysis}
Dulac-Arnold, G.; Levine, N.; Mankowitz, D.J.; Li, J.; Paduraru, C.; Gowal, S.;
  Hester, T.
\newblock {Challenges of real-world reinforcement learning: definitions,
  benchmarks and analysis}.
\newblock {\em Machine Learning} {\bf 2021}, {\em 110},~2419--2468.
\newblock
  doi:{\changeurlcolor{black}\href{https://doi.org/10.1007/S10994-021-05961-4/TABLES/11}{\detokenize{10.1007/S10994-021-05961-4/TABLES/11}}}.

\bibitem[Brunke \em{et~al.}(2021)Brunke, Greeff, Hall, Yuan, Zhou, Panerati,
  and Schoellig]{Brunke2021SafeLearning}
Brunke, L.; Greeff, M.; Hall, A.W.; Yuan, Z.; Zhou, S.; Panerati, J.;
  Schoellig, A.P.
\newblock {Safe Learning in Robotics: From Learning-Based Control to Safe
  Reinforcement Learning}.
\newblock {\em Annual Review of Control, Robotics, and Autonomous Systems} {\bf
  2021}, {\em 5}.
\newblock
  doi:{\changeurlcolor{black}\href{https://doi.org/10.1146/annurev-control-042920-020211}{\detokenize{10.1146/annurev-control-042920-020211}}}.

\bibitem[Mattsson \em{et~al.}(1998)Mattsson, Elmqvist, and
  Otter]{Mattsson1998PhysicalModelica}
Mattsson, S.E.; Elmqvist, H.; Otter, M.
\newblock {Physical system modeling with Modelica}.
\newblock  Control Engineering Practice. Pergamon,  1998, Vol.~6, pp. 501--510.
\newblock
  doi:{\changeurlcolor{black}\href{https://doi.org/10.1016/S0967-0661(98)00047-1}{\detokenize{10.1016/S0967-0661(98)00047-1}}}.

\bibitem[Gr{\"{a}}ber \em{et~al.}(2017)Gr{\"{a}}ber, Fritzsche, and
  Tegethoff]{Graber2017FromProblems}
Gr{\"{a}}ber, M.; Fritzsche, J.; Tegethoff, W.
\newblock {From system model to optimal control - A tool chain for the
  efficient solution of optimal control problems}.
\newblock  Proceedings of the 12th International Modelica Conference, Prague,
  Czech Republic, May 15-17, 2017. Link{\"{o}}ping University Electronic Press,
   2017, Vol. 132, pp. 249--254.
\newblock
  doi:{\changeurlcolor{black}\href{https://doi.org/10.3384/ecp17132249}{\detokenize{10.3384/ecp17132249}}}.

\bibitem[Brockman \em{et~al.}(2016)Brockman, Cheung, Pettersson, Schneider,
  Schulman, Tang, and Zaremba]{Brockman2016OpenAIGym}
Brockman, G.; Cheung, V.; Pettersson, L.; Schneider, J.; Schulman, J.; Tang,
  J.; Zaremba, W.
\newblock {OpenAI Gym}.
\newblock {\em arxiv} {\bf 2016}.

\bibitem[Lukianykhin and Bogodorova(2019)]{Lukianykhin2019ModelicaGym:Models}
Lukianykhin, O.; Bogodorova, T.
\newblock {ModelicaGym: Applying reinforcement learning to Modelica models}.
\newblock {\em ACM International Conference Proceeding Series} {\bf 2019}, pp.
  27--36.
\newblock
  doi:{\changeurlcolor{black}\href{https://doi.org/10.1145/3365984.3365985}{\detokenize{10.1145/3365984.3365985}}}.

\bibitem[Andersson \em{et~al.}(2016)Andersson, Akesson, and
  Fuhrer]{Andersson2016PyFMI:Interface}
Andersson, C.; Akesson, J.; Fuhrer, C.
\newblock {PyFMI: A Python Package for Simulation of Coupled Dynamic Models
  with the Functional Mock-up Interface}.
\newblock Technical Report~2, Lund University,  2016.

\bibitem[Raffin \em{et~al.}(2021)Raffin, Hill, Gleave, Kanervisto, Ernestus,
  and Dormann]{stable-baselines3}
Raffin, A.; Hill, A.; Gleave, A.; Kanervisto, A.; Ernestus, M.; Dormann, N.
\newblock {Stable-Baselines3: Reliable Reinforcement Learning Implementations}.
\newblock {\em Journal of Machine Learning Research} {\bf 2021}, {\em
  22},~1--8.

\bibitem[{OpenAI}(2020)]{OpenAI2020TwinDocumentation}
{OpenAI}.
\newblock {Twin Delayed DDPG — Spinning Up documentation},  2020.

\end{thebibliography}



\end{document}